\documentclass[aps,tightenlines,showpacs]{revtex4}
\usepackage{graphicx}
\usepackage{float}

\newcommand{\bea}{\begin{eqnarray}}
\newcommand{\eea}{\end{eqnarray}}
\newcommand{\xx}{\noindent}
\newcommand{\ra}{\rightarrow}

\begin{document}

\title{\Large{Real time statistical field theory}}

\author{ M.E. Carrington${}^{a,b}$, T. Fugleberg${}^{a,b}$, D.S. Irvine${}^c$, D. Pickering${}^c$.}
\email{carrington@brandonu.ca; fuglebergt@brandonu.ca; sammy@hotmail.com, pickering@brandonu.ca}
 \affiliation{ ${}^a$ Department of Physics, Brandon University, Brandon, Manitoba, R7A 6A9 Canada\\
 ${}^b$  Winnipeg Institute for Theoretical Physics, Winnipeg, Manitoba \\
 ${}^c$ Department of Mathematics, Brandon University, Brandon, Manitoba, R7A 6A9 }

\begin{abstract}
We have written a {\it Mathematica} program that calculates the
integrand corresponding to any amplitude in the closed-time-path
formulation of real time statistical field theory. 
The program is designed so that it can be used by someone with no 
previous experience with {\it Mathematica}. It
performs the contractions over the tensor indices that appear in real
time statistical field theory and gives the result in the 1-2,
Keldysh or RA basis. We have used the program to calculate the ward
identity for the QED 3-point function, the QED 4-point function
for two photons and two fermions, and the QED 5-point function
for three photons and two fermions. In real time statistical
field theory, there are seven 3-point functions, 15
 4-point functions and 31 5-point functions. We produce a table that gives the
results for all of these functions. In addition, we give a simple
general expression for the KMS conditions between $n$-point green
functions and vertex functions, in both the Keldysh and RA bases.
\end{abstract}


\date{\today}

\maketitle

\section{Introduction}

It is well known that calculations in real time statistical field theory are complicated by the proliferation of indices that results from the doubling of degrees of freedom (for a recent review see \cite{TonyRev}). This difficulty causes many people to avoid using real time finite temperature field theory, in spite of its significant advantages over the imaginary time formalism. Two of the major advantages of working in real time are that analytic continuations are not necessary, and that it  
is easy to generalize to non-equilibrium situations. 
In this paper we make a contribution towards reducing the technical difficulties associated with the real time formulation of statistical field theory. 

We use the closed time
path (CTP) formalism of real time statistical field theory \cite{Sch,Keld} which consists of a contour with two branches:  one 
runs from minus infinity to infinity along the real axis, the other
runs back from infinity to minus infinity just below the real
axis. This contour results in a doubling of degrees of freedom.  Physically, these extra
contributions come from the additional processes that are present when
the system interacts with a medium, instead of sitting in a vacuum.
As a result of these extra degrees of freedom, $n$-point functions
have a tensor stucture  which results in
calculational complexities that increase geometrically when one
considers either calculations at higher loop order, or the calculation
of higher $n$-point functions.  Statistical field theory can be
formulated in different bases, which produce different representations
of these tensors. There are three popular bases: the 1-2 basis, the
Keldysh basis, and the RA basis. Most of the original work in this
field was done in the 1-2 basis.  The Keldysh basis has the advantage
of being more easily adaptible to non-equilibrium situations. The RA basis 
produces particularly simple expressions in equilibrium.

This paper is organized as follows: In section \ref{1-2} we review the
1-2 formalism. In section \ref{basis} we discuss the issue of basis
transformations in general. Our discussion follows that of \cite{Gelis}. 
In sections \ref{Keldysh} and \ref{RA} we
give expressions for $n$-point functions and vertex functions in the
Keldysh and RA bases. In section \ref{KMS} we give the KMS relations
in each of these bases. These equations give a set of relations
between the various components of a given $n$-point function or vertex
function which hold in equilibrium, and are often useful for
simplifying the expressions that result after contracting over
indices. We note that some of these expressions have appeared
previously in the literature, using a slightly different notation \cite{mc-KMS,Heinz-Wang}. We give general expressions in the Keldysh and RA bases. We emphasize the
simplicity of the expressions in the RA basis. In section
\ref{program} we discuss the {\it Mathematica} computer program that 
we use to perform most of the calculations in this paper. This
program is designed so that it can be used by someone with no previous experience with {\it Mathematica}. It is available on the internet at
www.brandonu.ca/physics/fugleberg/Research/Dick.html.  

The program performs contractions over CTP indices and produces the integrand that corresponds to any $n$-point vertex
diagram, in the 1-2, Keldysh or RA basis. Several options to simplify the
results are built into the program and can be selected by the user as
part of the input. For example, for calculations in equilibrium, the
KMS conditions can be automatically implemented. We discuss the basic
design of the program and explain in detail, using a simple example,
how the program can be used.  
The program  
treats all fields as scalar bosons and 
from the beginning of the paper to the end of section \ref{program} we discuss only scalar bosons.
It is straightforward to apply the results of the program to other field theories. 
In section
\ref{WI} we illustrate the use of the program with the calculation of 
some QED ward identities. The general form is the same as the
well known zero temperature expression. In real time statistical field
theory however, the situation is complicated by the additional degrees
of freedom. Each $n$-point function has $2^n-1$ independent components
and each of these components has its own ward identity. We start by looking
at bare 1-loop diagrams and obtaining the form of the ward identities for
the  3-point function ($2^n-1 = 7$ components), the 4-point function for
two fermions and two photons (($2^n-1 = 15$ components), and the 5-point
function for two fermions and three photons ($2^n-1 = 31$ components). 
For the 3- and 4-point functions, we verify that the same expressions hold when full corrected vertices and propagators are used. 
Results for
some components of the 3-point function were obtained in \cite{mc-neq} and \cite{mc1}. A partial set of
expressions for bare 1-loop 3- and 4-point diagrams
in the htl approximation was derived in \cite{mc-spectral}. In this paper we produce a complete set of expressions. Our results agree with the partial results of \cite{mc-neq,mc1,mc-spectral}, which provides a check of our program. 
We note that the full
calculation involving corrected vertices would be prohibitatively
difficult without the use of a program like the one presented here. Our conclusions are presented in section \ref{conc}. 

\section{The 1-2 Basis}
\label{1-2}

The $n$-point function is defined:
\begin{equation}
G^{(n)}(x_1,\cdots x_n)_{b_1 \cdots b_n}:= (-i)^{n-1} \langle {\cal P}
\left(\phi(x_1)_{b_1} \cdots \phi(x_n)_{b_n}\right)\rangle
\end{equation}
where the subscripts $\{b_i\}$ take values 1 or 2 to indicate which
branch of the contour the corresponding time argument falls on, and
the symbol ${\cal P}$ represents ordering along the closed time
path. In what follows we will suppress the superscript ${(n)}$ and let
the number of indices indicate the number of fields in the $n$-point
function. All $n$-point functions will be denoted $G,$ except for the
2-point function, or the propagator, which will be called $D$.

In the 1-2 basis the 2-point function can be written as a $2\times 2$ matrix:
\begin{equation}
{\mathbf D}_{(1-2)}=\left(
\begin{array}{lr}
D_{11} & D_{12} \\
D_{21} & D_{22}
\label{propagator}
\end{array}
 \right)\,.
\end{equation}
The component $D_{11}$ indicates a propagator for fields moving along
the top branch of the contour, $D_{12}$ is the propagator for fields
moving from the top branch to the bottom branch, etc.  In co-ordinate
space these components are given by,
\begin{eqnarray}
\label{D12}
D_{11}(x,y)&=&-i \langle{T}(\phi(x)\phi(y)) \rangle \\
D_{12}(x,y)&=&-i \langle\phi(y)\phi(x) \rangle \nonumber\\
D_{21}(x,y)&=&-i \langle\phi(x)\phi(y) \rangle \nonumber \\
D_{22}(x,y)&=&-i \langle\tilde{T}(\phi(x)\phi(y)) \rangle\,,\nonumber
\end{eqnarray}
where ${\cal T}$ is the usual time ordering operator and $\tilde{\cal
T}$ is the antichronological time ordering operator.  These four
components satisfy, 
\begin{equation} \label{circ2} \sum_{a=1}^2 \sum
_{b=1}^2 (-1)^{a+b}D_{ab} = D_{11} - D_{12} - D_{21} + D_{22} = 0
\end{equation}
 as a consequence of the identity $\theta(x) +\theta(-x) =1$, and thus only three 
components are independent.

The 3-point function in the 1-2 basis is a $(2\times 2\times 2)$ tensor with components in co-ordinate space:
 \begin{eqnarray}
   G_{111}(x,y,z) 
    &=& -\langle  {\cal T}(\phi(x)\phi(y)\phi(z))\rangle  \nonumber\\
   G_{112}(x,y,z) 
    &=& -\langle\phi(z)\,{\cal T}(\phi(x)\phi(y))\rangle  \nonumber\\ 
   G_{121}(x,y,z)
    &=& -\langle\phi(y)\,{\cal  T}(\phi(x)\phi(z))\rangle \nonumber\\
   G_{211}(x,y,z)
    &=& -\langle\phi(x)\,{\cal  T}(\phi(y)\phi(z))\rangle \nonumber\\
   G_{122}(x,y,z)
    &=& -\langle\tilde{\cal  T}(\phi(y)\phi(z))\,\phi(x)\rangle \nonumber\\
   G_{212}(x,y,z) 
    &=& -\langle\tilde{\cal  T}(\phi(x)\phi(z))\,\phi(y)\rangle \nonumber\\
   G_{221}(x,y,z)
    &=& -\langle\tilde{\cal T}(\phi(x)\phi(y))\,\phi(z)\rangle \nonumber\\
   G_{222}(x,y,z)
    &=& -\langle\tilde{\cal T}(\phi(x)\phi(y)\phi(z))\rangle\, .
 \label{compVer}
 \end{eqnarray}
Only seven of these components are independent because of the identity
 \begin{equation} 
   \sum_{a=1}^2\sum_{b=1}^2\sum_{c=1}^2 
   (-1)^{a+b+c+1} G_{abc} =0 
 \label{circ3}
 \end{equation}
which follows in the same way as~(\ref{circ2}) from $\theta(x) + \theta(-x) = 1$.

The 4-point function in the 1-2 basis is a $(2\times 2\times 2\times 2)$ tensor. 
We write out a few examples:
\bea
&&G_{1111}(x,y,z,w) = i\langle {\cal T}\phi(x)\phi(y)\phi(z)\phi(w)\rangle\nonumber \\
&&G_{1112}(x,y,z,w) = i\langle \phi(w) {\cal T}(\phi(x)\phi(y)\phi(z))\rangle \nonumber \\
&&G_{1121}(x,y,z,w) = i\langle \phi(z) {\cal T}(\phi(x)\phi(y)\phi(w))\rangle \nonumber \\
&&G_{1211}(x,y,z,w) = i\langle \phi(y) {\cal T}(\phi(x)\phi(z)\phi(w))\rangle \nonumber \\
&&G_{2111}(x,y,z,w) = i\langle \phi(x) {\cal T}(\phi(y)\phi(z)\phi(w))\rangle \nonumber \\
&&G_{1122}(x,y,z,w) = i\langle \tilde {\cal T}(\phi(z)\phi(w)) {\cal T}(\phi(x)\phi(y))\rangle \nonumber \\
&& ~~~~~~~~~~~~~~~~~~~~~~~\vdots
\eea
These functions obey a relation similar to (\ref{circ2}) and (\ref{circ3}):
\bea
\label{circ4}
 \sum_{a=1}^2\sum_{b=1}^2\sum_{c=1}^2 \sum_{d=1}^2
   (-1)^{a+b+c+d} G_{abcd} =0 \,.
 \eea
The structure of higher $n$-point functions is similar.

Truncated green functions are called vertices and will be denoted $\Gamma$, except for the two point function which will be called $\Pi$. They are defined by the equation: 
\begin{equation}
\label{trun}
G_{b_1\cdots b_n}= G_{b_1 \bar b_1} \cdots G_{b_n \bar b_n} 
	\Gamma^{\bar b_1\cdots \bar b_n}  \,.
\end{equation}
The vertex functions satisfy the constraint:
\bea
\sum_{b_1=1}^2 \sum_{b_2=1}^2 \cdots \sum_{b_n=1}^2 \Gamma^{b_1\,b_2 \cdots b_n} = 0
\eea

\section{Basis transformations}
\label{basis}

A scattering amplitude is calculated by multiplying together a series
of vertices and propagators. By convention we will assign lower
indices to $n$-point functions (like the propagator) and upper indices
to vertices.  Summations are carried out over pairs of repeated
indices, where one of the indices is an upper index and the other is a
lower index.

The expression for a given scattering amplitude can be transformed to
a different basis by performing a rotation. There are two commonly used bases: the Keldysh basis and the R/A basis. 
Both of these bases express results in terms of combinations of the components of the 1-2 propagator that have a direct physical interpretation: the retarded, advanced and symmetric propagators. These expressions are easy to obtain in co-ordinate
space. Using (\ref{D12}) and (\ref{circ2}) one can show:
\begin{eqnarray} 
\label{3a}
&&  D_{ret}(x,y) = D_{11}(x,y) - D_{12}(x,y)  = -i\theta(x_0-y_0)\langle [\phi(x),\phi(y)]\rangle  
	 \\
&&  D_{adv}(x,y)  = D_{11}(x,y) - D_{21}(x,y) = -i\theta(y_0-x_0)\langle [\phi(x),\phi(y)] \rangle 
	 \nonumber\\ 
&&  D_{sym}(x,y)  = D_{11}(x,y) + D_{22}(x,y) = -i\langle \{\phi(x),\phi(y) \rangle 
	\, .\nonumber
\end{eqnarray}
One advantage of the
Keldysh basis is that it is easily generalizable to non-equilibrium situations. 
In equilibrium, the R/A basis produces particularly simple expressions for amplitudes, and for expressions
like the KMS conditions, which give relationships between different
amplitudes.

\begin{figure}[H]
\begin{center}
\includegraphics[width=8cm]{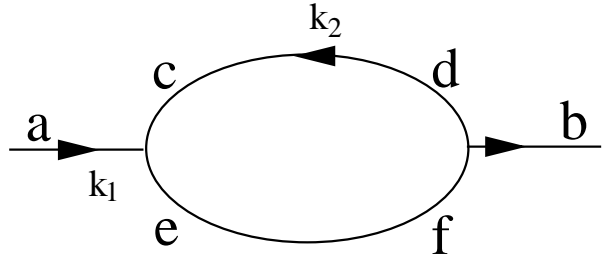}
\end{center}
\caption{A typical ampitude}
 \label{KMS1}
\end{figure}

The rotation to a different basis is
accomplished by matrix multiplication. We rotate lower indices by
multiplying by a matrix $U$ and upper indices by multiplying by the
matrix $V$.   These matrices are related through the following equation.
\bea
\label{Veqn}
V(k) = (U^{T})^{-1}(-k)\,.
\eea 
We obtain expressions of the form:
\bea
\label{trans1}
&&U_{\bar a}^{~ a}(k) U_{\bar b}^{~ b}(-k)D_{ab}(k) \to  D^\prime_{\bar a,\bar b}(k) \\
&&V^{\bar a}_{~ a}(k_1)
	V^{\bar b}_{~ b}(k_2) V^{\bar c}_{~ c}(-k_1-k_2)\Gamma^{abc}(k_1,k_2,-k_1-k_2)\to (\Gamma^\prime)^{\bar a,\bar b,\bar c}(k_1,k_2,-k_1-k_2)  \nonumber
\eea
where we have simplified the notation by writing a two point function of the form $D_{ab}(k,-k)$ as $D_{ab}(k)$.
It is straightforward to see that amplitudes have the correct transformation properties. 
We look at the example shown in Fig. \ref{KMS1}. 
\par
\noindent The amplitude is represented by an expression of the form:
\bea
\Pi^{ab}(k_1) \sim \int dk_2\;\Gamma^{ace}(k_1,k_2,-k_1-k_2)D_{cd}(-k_2)
\Gamma^{bdf}(-k_1,-k_2,k_1+k_2)D_{ef}(k_1+k_2)
\eea
Using (\ref{Veqn}) and (\ref{trans1}) it is easy to see that this self energy transforms 
according to:
\bea
\Pi^{ab}(k_1) = V^a_{~~\bar a}(k_1)V^b_{~~\bar b}(-k_1)\Pi^{\bar a \bar b}(k_1)\,.
\eea

In the sections below, indices in the 1-2 basis will
be denoted $b_i$ and take the values 1 or 2.  Keldysh indices will
be written $\alpha_i$ and are assigned the values $\alpha = 1 := r$
and $\alpha = 2:= a$. R/A indices will be denoted $X_i$ and assigned
the values $X = 1:=R$ and $X=2:=A$.  An $n$-point function in the 1-2
basis will be written:
\bea
G(k_1,k_2,k_3,\cdots k_n)_{b_1,b_2,b_3,\cdots b_n} := G_{b_1,b_2,b_3,\cdots b_n}
\eea
An $n$-point function in the Keldysh basis will be written:
\bea
G(k_1,k_2,k_3,\cdots k_n)_{\alpha_1,\alpha_2,\alpha_3,\cdots \alpha_n} := 
	G_{\alpha_1,\alpha_2,\alpha_3,\cdots \alpha_n}
\eea
An $n$-point function in the R/A basis will be written:
\bea
G(k_1,k_2,k_3,\cdots k_n)_{X_1,X_2,X_3,\cdots X_n} := G_{X_1,X_2,X_3,\cdots X_n} 
\eea
Note that the conservation of momentum gives $k_n = -(k_1+k_2+\cdots +k_{n-1})$ so that each $n$-point function depends on only $n-1$ independent momenta. 

\section{Thermal Functions}

We write the general distribution function as $f(k)$ and define the symmetric and anti-symmetric combinations:
\begin{eqnarray}
&&{\cal F}_A(k)=f(k)-f(-k)\,;~~{\cal F}_S(k)=f(k)+f(-k)
\eea
In the Keldysh representation it will be useful to define the functions:
\bea
\tilde {\cal F}(k_i;\alpha_i) = \delta_{\alpha_i,a}{\cal F}_A(k_i) - \delta_{\alpha_i,r}{\cal F}_S(k_i)\,;~~ \hat {\cal F}(k_i;\alpha_i) = \delta_{\alpha_i,r} {\cal F}_A(k_i)
	- \delta_{\alpha_i,a}{\cal F}_S(k_i).
\eea
To illustrate these definitions consider the expression:
\bea
\tilde {\cal F}(k_1;\alpha_1)\tilde {\cal F}(k_2;\alpha_2)\tilde {\cal F}(k_3;\alpha_3)G(k_1,k_2,k_3)_{\alpha_1,\alpha_2,\alpha_3}
\eea
If we take the case $\alpha_1 = \alpha_2 = r\,;~\alpha_3 = a$ we obtain:
\bea
{\cal F}_S(k_1) {\cal F}_S(k_2){\cal F}_A(k_3) G(k_1,k_2,k_3)_{rra}.
\eea
For bosons in equilibrium these expressions become:
\bea
\label{equib1}
&& f(k) \rightarrow n(k) = 1/(e^{\beta k_0}-1)\\
&& \tilde {\cal F}(k_i;\alpha_i)\rightarrow \tilde N(k_i;\alpha_i)\,;~~
	\hat {\cal F}(k_i;\alpha_i)\rightarrow \hat N(k_i;\alpha_i) \nonumber\\
&& {\cal F}_S(k_i)\rightarrow -1\,;~~ {\cal F}_A(k_i)\rightarrow N(k_i) \nonumber
\eea
where we have defined:
\bea
\label{therm-basic}
&& \tilde N(k_i;\alpha_i) = \delta_{\alpha_i,a}N(k_i) + \delta_{\alpha_i,r}\\
&& \hat N(k_i;\alpha_i) = \delta_{\alpha_i,r}N(k_i) + \delta_{\alpha_i,a}.\nonumber \\
&& N_i:=N(k_i) = 1+2n(k_i)\nonumber
\eea
Equilibrium functions satisfy:
\bea
\label{therm-prop}
&& N_1+N_2 = 0\,\mbox{ if } k_2=-k_1 \\
&& 1+N_1 N_2+N_1 N_3+N_2 N_3 = 0\,\mbox{ if } k_3=-k_1-k_2 \nonumber\\
&& N_1+N_2+N_3+N_4+N_1 N_2 N_3+N_2 N_3 N_4+ N_3 N_4 N_1+N_4 N_3 N_2 = 0\,\mbox{ if } k_4=-k_1-k_2-k_3\nonumber\\
&&~~~~~~~~~~~~\vdots\nonumber
\eea
where the dots indicate that higher order expressions can be generated
by iteration. (Expressions with the same properties can be defined for
the fermion distribution function: $ n_f(k)= 1/(e^{\beta
k_0}+1)\,;~~N_F(k_i) = 1-2n_f(k_i) $).

In order to write the KMS conditions in
a concise way we define the function
\bea
C_n : = C(\{N_i\}) = C(N_1,N_2,\cdots N_n) = \sum^{n}_{p=0} \frac{1}{2}\Big[1-(-1)^{n+p}\Big]\;\cdot
\;{\bf N}(n;p) 
\label{C_definition}
\eea
where the symbol ${\bf N}(n;p)$ means the following:

\xx [a] start with $n$ indices $\{x_1,x_2,x_3\,\cdots x_n\}$

\xx [b] consider all possible subsets of these indices containing $p<n$ of the $x_i$'s (without considering order)

\xx [c] for each of these subsets, take the product of the corresponding  $N(x_i)$'s

\xx [d] sum over all sets

\xx In addition, we define ${\bf N}(n;0) = 1$.
A few examples will illustrate this notation. \\

\xx Example [1]: if $n=3$ and $p=2$ then the possible sets of $p$ are: $\{x_1,x_2\}$, $\{x_2,x_3\}$, $\{x_3,x_1\}$ and the result is ${\bf N}(3;2) = N_1 N_2+N_2 N_3+N_3 N_1$.\\

\xx Example [2]: if $n=4$ and $p=3$ then the possible sets of $p$ are: $\{x_1,x_2,x_3\}$, $\{x_2,x_3,x_4\}$, $\{x_3,x_4,x_1\}$, $\{x_4,x_1,x_2\}$ and the result is ${\bf N}(4;3) = N_1 N_2 N_3 +N_2 N_3 N_4 +N_3 N_4 N_1+N_4 N_1 N_2$.\\

\xx Example [3]: if $n=3$ and $p=1$ then the possible sets of $p$ are: $\{x_1\}$, $\{x_2\}$, $\{x_3\}$ and the result is ${\bf N}(3;1) = N_1+N_2+N_3$.\\

\xx Below we write out the first few  $C(N_1,N_2,\cdots N_n)$'s:
\bea
&& C(N_1) = 1\\
&&C(N_1,N_2)= N_1+N_2\nonumber \\
&&C(N_1,N_2,N_3)   = 1+N_1 N_2+N_2 N_3+ N_3 N_1\nonumber\\
&&C(N_1,N_2,N_3,N_4) = N_1+N_2+N_3+N_4+N_1 N_2 N_3+N_2 N_3 N_4+ N_3 N_4 N_1+N_4 N_3 N_2\nonumber
\eea

\noindent Note that because of (\ref{therm-prop}) each of these expressions is zero if the momenta 
satisfy $k_1+k_2+\cdots k_n =0$.
Using (\ref{therm-basic}) and (\ref{therm-prop}) we have:
\begin{equation}
C(N_1,\cdots N_n)={2^{n-1}} \frac{n(k_1)\cdots n(k_n)}{n\left(k_1+\cdots + k_n\right) }\,.
\label{Cintermsofns}
\end{equation}

To write the KMS conditions for a given $n$-point function in a
compact form, we will need to use $C$-functions of the form defined
above, but with arguments that are not the full set $\{N_i\}$
with $i$ running from 1 to $n$. We will need the $C$ variables whose
arguments are a subset of the $N_i$'s. When working in the Keldysh
basis we want $C$-functions whose arguments are the set of $N_i$'s
whose corresponding $\alpha_i$'s take the value [i] $r$ or; [ii]
$a$. When working in the R/A basis we want $C$-functions whose
arguments are the set of $N_i$'s whose the corresponding $X_i$'s take
the value [i] $R$ or; [ii] $A$.\\

\xx We define these modified $C$-functions as follows. In the Keldysh basis the set $\{\alpha_i\}$ contains 
$n$ variables. The number of $r$'s is defined to be $n_r$ and the
number of $a$'s is defined to be $n_a$.  Of course, we have $n =
n_r+n_a$. We construct a set of $n_r$ variables: $\{N_i
\delta_{\alpha_i,r}\}$ and a set of $n_a$ variables: $\{N_i
\delta_{\alpha_i,a}\}$. Using these sets of $N$'s we define the
corresponding $C$-functions:
\bea
&&C_{n_r} : =  C(\{N_i \delta_{\alpha_i,r}\}) \\
&&C_{n_a} : = C(\{N_i \delta_{\alpha_i,a}\})\,. \nonumber
\eea
\xx When working in the R/A basis we make the analogous definitions: The set $\{X_i\}$ 
contains $n$ variables. The number of $R$'s is $n_R$ and the number of $A$'s is $n_A$. 
We define:
\bea
&&C_{n_R}:= C(\{N_i \delta_{X_i,R}\})\\ &&C_{n_A}:= C(\{N_i
\delta_{X_i,A}\})\,.\nonumber
\eea
To clarify this notation, we look at the following example: Take $n=7$ and consider the set $\{\alpha_i\}$  = $\{r,r,a,r,a,a,r\}$. 
We have: $n_r$ = 4; $C_{n_r} = C(N_1,N_2,N_4,N_7)$ and $n_a = 3$; $C_{n_a} = C(N_3,N_5,N_6)$.

\section{The Keldysh Basis}
\label{Keldysh}
The rotation from the 1-2 representation to the Keldysh representation 
is accomplished by using the transformation matrix:
\begin{equation}
\label{firstU}
U_{Keldysh\leftarrow (1-2) }=\frac{1}{\sqrt{2}}\left(
\begin{array}{lr}

1 & 1 \\
1 & -1
\end{array}
\right).
\end{equation}
Note that in this case (\ref{Veqn}) gives $V=U$.  In general the $n$-point
function in the Keldysh representation is given by:
\begin{equation}
\label{Keldyn}
G_{\alpha_1\cdots\alpha_n}=2^{\frac{n}{2}-1}\, U_{\alpha_1}\!^{b_1} 
\cdots U_{\alpha_n}\!^{b_n} G_{b_1\cdots b_n}\,.
\end{equation}
The factor $2^{n/2-1}$ is introduced to produce simpler expressions for the vertices, following  \cite{Heinz-Wang} and \cite {Chou}.
For example, using the notation defined in (\ref{Keldyn}), the bare 3-point vertex in the Keldysh representation is written in tensor form:
\bea
\{
 \{0,-i\} , \{-i,0\} ,
 \{-i,0\} , \{0,-i\}\}
\eea
This expression can be compared with the bare 3-point vertex in the 1-2 representation:
\bea
\{
 \{-i,0\} , \{0,0\} ,
 \{0,0\} , \{0,i\}\}
\eea

In addition, the appropriate components of the Keldysh vertex correspond directly to the retarded vertices
(see for example Eqn. (\ref{fac-exp})).

\subsection{2-point Function}

The 2-point function in the Keldysh basis is obtained from (\ref{Keldyn}):
\begin{equation}
D_{\alpha_1\alpha_2}= U_{\alpha_1}\!^{b_1}  U_{\alpha_2}\!^{b_2} 
G_{b_1 b_2}
\end{equation}
and can be represented (using (\ref{circ2}) and (\ref{3a})) by:
\begin{eqnarray}
\label{keld-prop}
{\bf D}_{Keldysh}&:=&
\left(
\begin{array}{lr}
D_{rr} & D_{ra}\\
D_{ar} & D_{aa}
\end{array} 
\right)\\
&=& U\cdot {\bf D}_{(1-2)}\cdot U^T= \left(
\begin{array}{lc}
D_{11}+D_{22} & D_{11}-D_{12}\\
D_{11}-D_{21} & 0
\end{array}
\right)
= \left(
\begin{array}{lc}
D_{sym} & D_{ret}\\
D_{adv} & 0
\end{array}
\right) \,.\nonumber
\end{eqnarray}
Note that $D_{aa}=0$. In fact, for any $n$-point function in the
Keldysh basis, it is always true that
\bea 
G_{aaa\cdots a}=0\,.
\label{aaaa} 
\eea 
To understand this point, recall that according to
(\ref{circ2}), (\ref{circ3}) and (\ref{circ4}) it is always possible
to express one component of the tensor 2-, 3- and 4-point functions in
terms of the other components. It is straightforward to show that in
the 1-2 basis a constraint of the same form as (\ref{circ2}),
(\ref{circ3}) and (\ref{circ4}) exists for any arbitrary $n-$point
function. In the Keldysh representation this constraint takes the form
(\ref{aaaa}). The vertex function obeys an analogous constraint:
\bea 
\Gamma^{rrr\cdots r}=0\,.
\label{rrrr} 
\eea \\
 
 \subsection{3-point Function}
 The 3-point function in the Keldysh representation is obtained from (\ref{Keldyn}):
 \bea
 G_{\alpha_1\alpha_2\alpha_3} = \sqrt{2} \, U_{\alpha_1}^{~~b_1}U_{\alpha_2}^{~~b_2}U_{\alpha_3}^{~~b_3}G_{b_1 b_2 b_3}\,.
 \eea
We give three examples of the results that we obtain by summing over indices and using (\ref{circ3}):
 \bea
 \label{3-ex}
 &&G_{rrr} = G_{111}+G_{221}+G_{212}+G_{122}\\
 &&G_{rra} = G_{111}-G_{112}+G_{221}-G_{222}\nonumber\\
 &&G_{raa} = G_{111}-G_{112}-G_{121}+G_{122}\,.\nonumber
 \eea
Using (\ref{compVer}) it is easy to show in co-ordinate space that 
 \bea
 \label{fac-exp}
 G_{raa}(x_1,x_2,x_3) = G_{R_1}(x_1,x_2,x_3)
 \eea
where $G_{R_1}$ is the component of the 3-point function that is retarded with respect to the first leg and 
is given by the familiar expression:
 \bea
 && G_{R_1}(x_1,x_2,x_3)\\
 &&~~= \left(-i\right)^2\Big(\theta(t_1-t_2)\theta(t_2-t_3)\langle [[\phi(x_1),\phi(x_2)],\phi(x_3)]\rangle + 
\theta(t_1-t_3)\theta(t_3-t_2)\langle [[\phi(x_1),\phi(x_3)],\phi(x_2)]\rangle\Big)\,\nonumber\\
  &&~~=  \left(-i\right)^2\sum_{\{t_a,t_b\}=\{t_2,t_3\}}\theta(t_1,t_a)\theta(t_a,t_b)\langle [[\phi(t_1),\phi(t_a)],\phi(t_b)]\rangle\,.\nonumber
 \eea
The summation indicates that we sum over the combinations: $(t_a = t_2;\;t_b = t_3)$ 
and $(t_a = t_3;\;t_b = t_2)$.  The other two components given in (\ref{3-ex}) correspond to similar expressions
in co-ordinate space, but involve anti-commutators.
 
\subsection{4-point function}
 
The 4-point function in the Keldysh representation is obtained from (\ref{Keldyn}):
\bea
 G_{\alpha_1\alpha_2\alpha_3\alpha_4} = 2 U_{\alpha_1}^{~~b_1}U_{\alpha_2}^{~~b_2}U_{\alpha_3}^{~~b_3}
U_{\alpha_4}^{~~b_4}G_{b_1 b_2 b_3 b_4}
\eea
We give one example of the results that we obtain by summing over indices and using (\ref{circ4}):
 \bea
&& G_{raaa} = G_{1111}-G_{1112}-G_{1121}-G_{1211}+G_{1122}+G_{1212}+G_{1221}-G_{1222}
\eea
In co-ordinate space this expression has the form:
\bea
&&G_{raaa}(x_1,x_2,x_3,x_4) =G_{R1}(x_1,x_2,x_3,x_4) \\ 
&&~~ = \left(-i\right)^3 \sum_{\{t_a,t_b,t_c\}=\{t_2,t_3,t_4\}}\theta(t_1,t_a)\theta(t_a,t_b)
\theta(t_b,t_c)\langle [[[\phi(t_1),\phi(t_a)],\phi(t_b)],\phi(t_c)]\rangle\nonumber 
 \eea
where the sum is over all possible assignments of the variables
$\{t_2,\,t_3,\,t_4\}$ to the variables $\{t_a,\,t_b,\,t_c\}$ and the
notation $G_{R1}(x_1,x_2,x_3,x_4)$ indicates the 4-point function that
is retarded with respect to the first leg. Expressions for higher
$n$-point functions are obtained similarly.

\section{The (R/A) Basis}
\label{RA}

The matrix that performs rotations from the 1-2 basis to the R/A basis is:
\begin{equation}
\label{12RA}
U_{(R/A)\leftarrow (1-2)}=
\left( 
\begin{array}{cc}
-f(k) & f(k) \\
1         & f(k)/ f(-k)
\end{array}  
\right)\,.
\end{equation}
We will find it more useful to obtain expressions in the RA basis by rotating from the Keldysh basis. 
The matrix that transforms from the Keldysh basis to the 1-2 basis is given by the 
inverse of (\ref{firstU}).
Combining with (\ref{12RA}) we obtain the matrix that transforms from the Keldysh basis to the R/A basis:
\bea
U_{(R/A) \leftarrow Keldysh } &&= U_{(R/A)\leftarrow (1-2)}(U_{Keldysh \leftarrow (1-2)})^{-1} \\
&& =  
\left(
\begin{array}{cc}
0 & -\sqrt{2} \, f(k) \\
\frac{{\cal F}_S(-k)}{\sqrt{2}\, f(-k)} & \frac{{\cal F}_A(-k)}{\sqrt{2}\, f(-k)}
\end{array}
\right)\,.\nonumber
\eea
The $n$-point function in R/A basis is given by:
\begin{equation}
2^{\frac{n}{2}-1}G_{X_1\cdots X_n}= U(k_1)_{X_1}^{~ \alpha_1} \cdots    U(k_n)_{X_n}^{~ \alpha_n} 
G_{\alpha_1\cdots \alpha_n} \,.
\label{RAdefn}
\end{equation}
General expressions that relate R/A $n$-point 
functions and vertex functions to Keldysh functions are given below:
\begin{eqnarray}
\label{RAone}
&&G(k_1,\cdots k_n)_{X_1\cdots X_n } \\
&&~~~~=\frac{(-1)^n}{2^{n-1}} \frac{\prod_{i}[\delta_{X_i,A}+\delta_{X_i,R}\;2 \, f(k_i)]}
{\prod_{i}[\delta_{X_i,R}+\delta_{X_i,A} \; f(-k_i)]} \cdot \sum_{\alpha_l\in\{r,a\}}
 \left( \prod_{l=1}^{n}  \left[\tilde{{\cal F}}(k_l;\alpha_l)\delta_{X_l A}+ \delta_{X_l R}
\delta_{\alpha_l a}\right] \right) G(k_1,\cdots k_n)_{   \alpha_{1} \cdots \alpha_n  } 
\nonumber\\
&&\Gamma(k_1,\cdots k_n)^{X_1\cdots X_n } \nonumber\\
&&~~~~=\frac{(-1)^n}{2^{n-1}} \frac{\prod_{i}[\delta_{X_i,R}+\delta_{X_i,A}\;2 \, f(k_i)]}
{\prod_{i}[\delta_{X_i,A}+\delta_{X_i,R} \; f(-k_i)]\left(-{\cal F}_S(k_i)\right)} \cdot \sum_{\alpha_l\in\{r,a\}}
 \left( \prod_{l=1}^{n}  \left[\hat{{\cal F}}(k_l;\alpha_l)\delta_{X_l R}+ \delta_{X_l A}
\delta_{\alpha_l r}\right] \right) \Gamma(k_1,\cdots k_n)^{   \alpha_{1} \cdots \alpha_n  } 
\nonumber
\eea
From (\ref{aaaa}), (\ref{rrrr}) and (\ref{RAone}) we have
\bea
G_{R\cdots R} = 0\,;~~\Gamma^{A\cdots A}=0\,.
\eea

These equations can be written in a simpler form in equilibrium. There are two kinds of simplifications that occur. We can replace general distribution functions with thermal distribution functions and use identities of the form (\ref{equib1}), (\ref{therm-basic}), (\ref{therm-prop}), (\ref{Cintermsofns}) for combining groups of distributions functions. We can also use KMS conditions. The KMS conditions are a set of equations which are valid only at equilibrium that relate various components of an $n$-point function or vertex function. These conditions will be derived in the next section. For completeness, we give below two general expressions: the first corresponds to (\ref{RAone}) using thermal distribution functons, and the second comes from (\ref{RAone}) using thermal distribution functions and the KMS conditions. 
The general equation obtained from (\ref{RAone}) using thermal distribution functions is:
\begin{eqnarray}
\label{RAonenewest}
&&{2^{n_R-1}} {(-1)^{n_R+1}}\;{C_{n_A}(\{N_i\delta_{X_i,A}\})}  G(k_1,\cdots k_n)_{X_1\cdots X_n } \\
&&~~~~= C_{n_R}(\{N_i\delta_{X_i,R}\}) \cdot \sum_{\alpha_l\in\{r,a\}}
 \left( \prod_{l=1}^{n}  \left[\tilde N(k_l;\alpha_{l})\delta_{X_l A}+ \delta_{X_l R}\delta_{\alpha_l a}\right] \right) G(k_1,\cdots k_n)_{   \alpha_{1} \cdots \alpha_n  }. \nonumber\\
&&{2^{n_A-1}} {(-1)^{n_A+1}}\;{C_{n_R}(\{N_i\delta_{X_i,R}\})}\Gamma(k_1,\cdots k_n)^{X_1\cdots X_n } \nonumber\\
&&~~~~= {C_{n_A}(\{N_i\delta_{X_i,A}\})}
 \cdot \sum_{\alpha_l\in\{r,a\}}
 \left( \prod_{l=1}^{n}  \left[\hat{N}(k_l;\alpha_l)\delta_{X_l R}+ \delta_{X_l A}
\delta_{\alpha_l r}\right] \right) \Gamma(k_1,\cdots k_n)^{   \alpha_{1} \cdots \alpha_n  }. 
\nonumber
 \eea 
After applying the KMS conditions to the right hand side of this expression we obtain:
\bea
\label{RA-keld-eq}
&&{2^{n_R-1}} {(-1)^{n_R+1}}\;  G(k_1,\cdots k_n)_{X_1\cdots X_n } \\
&&~~~~=  \sum_{\alpha_l\in\{r,a\}}
 \left( \prod_{l=1}^{n}  \left[\tilde N(k_l;\alpha_{l})\delta_{X_l R}+ \delta_{X_l A}\delta_{\alpha_l a}\right] \right) G^*(k_1,\cdots k_n)_{   \alpha_{1} \cdots \alpha_n  }. \nonumber
\eea
For clarity we give a few examples. We first give 
general expressions using (\ref{RAone}). We next give the results obtained with
thermal distribution functions using (\ref{RAonenewest}).  Finally, we give
the expressions obtained by applying the KMS conditions to the right hand
side of the previous equations, or equivalently from using (\ref{RA-keld-eq}).

\xx [i] The 2-point functions in the R/A formalism are given by:
\begin{eqnarray}
D_{RR}&=&0 \\
D_{RA}&=&-{\cal F}_S(k_2) D_{ar} \nonumber\\
D_{AR}&=&-{\cal F}_S(k_1)D_{ra} \nonumber\\ 
D_{AA}&=&  \left({\cal F}_S(k_1){\cal F}_S(k_1)D_{rr}-{\cal F}_A(k_1){\cal F}_S(k_2) D_{ar}-
{\cal F}_S(k_1){\cal F}_R(k_2) D_{ra}\right)/\left( 2 f(-k_1)f(-k_2)\right).\nonumber
\end{eqnarray}
Using equilibrium distribution functions these results reduce to:
\begin{eqnarray}
D_{RR}= 0\,;~~~~D_{RA}= D_{ar} \,;~~~~
D_{AR}= D_{ra}  \,;~~~~
D_{AA}= \frac{1}{2 n(-k_1)n(-k_2)}\left(D_{rr}(p) + N_1 D_{ar}+N_2 D_{ra}
\right).\nonumber
\end{eqnarray}
Applying KMS conditions we have:
\begin{eqnarray}
D_{RR}= 0\,;~~~~D_{RA}= D^*_{ra} \,;~~~~
D_{AR}= D^*_{ar}  \,;~~~~
D_{AA}= 0\,,\nonumber
\end{eqnarray}
which can be written in the familiar form:
\bea
{\bf D}_{R/A} = \left(\begin{array}{lr}
D_{RR} & D_{RA}\\
D_{AR} & D_{AA}\,,
\end{array} 
\right) = \left(\begin{array}{lr}
0 & D_{adv}\\
D_{ret} & 0
\end{array} 
\right)\,.
\eea

\xx [ii] The 3-point functions in the R/A formalism are given by:
\begin{eqnarray}
G_{RRA}&=& \frac{f(k_1)f(k_2)}{f(k_1+k_2)} {\cal F}_S(k_3)\; G_{aar} \label{RAintermsofKeldysh}\\
G_{RAR}&=& \frac{f(k_1)f(k_3)}{f(k_1+k_3)} {\cal F}_S(k_2)\; G_{ara}\nonumber \\
G_{ARR}&=& \frac{f(k_2)f(k_3)}{f(k_2+k_3)} {\cal F}_S(k_1)\; G_{raa}\nonumber \\
G_{AAR}&=&-\frac{1}{2} \frac{f(-k_1-k_2)}{f(-k_1)f(-k_2)}\left({\cal F}_S(k_1){\cal F}_S(k_2) 
		G_{rra} 
	- {\cal F}_A(k_1){\cal F}_S(k_2) G_{ara} - {\cal F}_S(k_1){\cal F}_A(k_2) 
		G_{raa} \right) \nonumber\\
G_{ARA}&=& -\frac{1}{2}\frac{f(-k_1-k_3)}{f(-k_1)f(-k_3)}\left({\cal F}_S(k_1){\cal F}_S(k_3) 
		G_{rar} 
	- {\cal F}_A(k_1){\cal F}_S(k_3) G_{aar} - {\cal F}_S(k_1){\cal F}_A(k_3)  
		G_{raa} \right) \nonumber\\
G_{RAA}&=& -\frac{1}{2}\frac{f(-k_2-k_3)}{f(-k_2)f(-k_3)}\left({\cal F}_S(k_2){\cal F}_S(k_3) 
		G_{arr} 
	- {\cal F}_A(k_2){\cal F}_S(k_3) G_{aar} - {\cal F}_S(k_2){\cal F}_A(k_3)  G_{ara} 
		\right)\nonumber\\
G_{AAA}&=& \frac{1}{4\; f(-k_1)\; f(-k_2) \; f(-k_3)} \nonumber\\[2mm]
~~~~~~~~&\cdot &\Big({\cal F}_S(k_1){\cal F}_S(k_2){\cal F}_S(k_3) G_{rrr} \nonumber\\
&&	-{\cal F}_A(k_1){\cal F}_S(k_2){\cal F}_S(k_3) G_{arr} 
	-{\cal F}_S(k_1){\cal F}_A(k_2){\cal F}_S(k_3) G_{rar} 
	-{\cal F}_S(k_1){\cal F}_S(k_2){\cal F}_A(k_3) G_{rra}\nonumber\\ 
&& 	+{\cal F}_A(k_1){\cal F}_A(k_2){\cal F}_A(k_3) G_{aar} 
	+{\cal F}_A(k_1){\cal F}_S(k_2){\cal F}_A(k_3) G_{ara} 
	+{\cal F}_S(k_1){\cal F}_A(k_2){\cal F}_A(k_3) G_{raa} \Big)\nonumber
\eea

Using equilibrium distribution functions these results reduce to:
\begin{eqnarray}
G_{RRA}&=&-\frac{(N_1+N_2)}{2} \; G_{aar} \label{RAintermsofKeldyshtwo}\\
G_{RAR}&=&-\frac{(N_1+N_3)}{2} \; G_{ara}\nonumber \\
G_{ARR}&=&-\frac{(N_2+N_3)}{2} \; G_{raa}\nonumber \\
G_{AAR}&=&\frac{1}{(N_1+N_2)} \left( G_{rra} + N_1 G_{ara} + N_2 G_{raa} \right) \nonumber\\
G_{ARA}&=&\frac{1}{(N_1+N_3)} \left( G_{rar} + N_1 G_{aar} + N_3 G_{raa} \right) \nonumber\\
G_{RAA}&=&\frac{1}{(N_2+N_3)} \left( G_{arr} + N_2 G_{aar} + N_3 G_{ara} \right) \nonumber\\
G_{AAA}&=&-\frac{1}{4\; n(-k_1)\; n(-k_2) \; n(-k_3)} \left( G_{rrr}+ N_1 G_{arr} + N_2 G_{rar} + N_3 G_{rra} + N_1 N_2 G_{aar} + N_1 N_3 G_{ara} + N_2 N_3 G_{raa} \right)\nonumber
\eea
Applying KMS conditions we have:
\begin{eqnarray}
G_{RRA}&=&-\frac{1}{2} \left(G^*_{rra} + N_1 G^*_{ara} + N^*_2 G_{raa}\right) 
\label{RAintermsofKeldyshthree}\\
G_{RAR}&=&-\frac{1}{2} \left(G^*_{rar} + N_1 G^*_{aar} + N_3 G^*_{raa} \right)
\nonumber \\
G_{ARR}&=&-\frac{1}{2} \left(G^*_{arr} + N_2 G^*_{aar} + N_3 G^*_{ara}\right)
\nonumber \\
G_{AAR}&=&G^*_{aar} \nonumber\\
G_{ARA}&=&G^*_{ara} \nonumber\\
G_{RAA}&=&G^*_{raa} \nonumber\\
G_{AAA}&=&0\,.\nonumber
\eea

For the 4-point functions using equilibrium distribution functions we have:
\begin{eqnarray}
G_{RRRA}&=&\frac{C(N_1,N_2,N_3)}{4} \; G_{aaar} \label{RAintermsofKeldyshtwo4ptone}\\
G_{RRAR}&=&\frac{C(N_1,N_2,N_4)}{4} \; G_{aara}\nonumber \\
G_{RARR}&=&\frac{C(N_1,N_3,N_4)}{4} \; G_{araa}\nonumber \\
G_{ARRR}&=&\frac{C(N_2,N_3,N_4)}{4} \; G_{raaa}\nonumber \\
G_{RRAA}&=&-\frac{1}{2}\frac{C(N_1,N_2)}{C(N_3,N_4)}
    \left( G_{aarr} \!\!+\!\! N_3 G_{aaar} \!\!+\!\! N_4 G_{aara} \right) \nonumber\\
G_{RARA}&=&-\frac{1}{2}\frac{C(N_1,N_3)}{C(N_2,N_4)}
    \left(  G_{arar} \!\!+\!\! N_2 G_{aaar} \!\!+\!\! N_4 G_{araa} \right) \nonumber\\
G_{RAAR}&=&-\frac{1}{2}\frac{C(N_1,N_4)}{C(N_2,N_3)}
    \left(G_{arra} \!\!+\!\! N_2 G_{aara} \!\!+\!\! N_3 G_{araa}  \right) \nonumber\\
G_{ARRA}&=&-\frac{1}{2}\frac{C(N_2,N_3)}{C(N_1,N_4)}
     \left( G_{raar} \!\!+\!\! N_1 G_{aaar} \!\!+\!\! N_4 G_{raaa} \right) \nonumber\\
G_{ARAR}&=&-\frac{1}{2}\frac{C(N_2,N_4)}{C(N_1,N_3)}
     \left(G_{rara} \!\!+\!\! N_1 G_{aara} \!\!+\!\! N_3 G_{raaa}  \right) \nonumber\\
G_{AARR}&=&-\frac{1}{2}\frac{C(N_3,N_4)}{C(N_1,N_2)}
      \left(G_{rraa} \!\!+\!\! N_1 G_{araa} \!\!+\!\! N_2 G_{raaa}  \right)\nonumber\\
G_{AAAR}&=&\frac{1}{C(N_1,N_2,N_3)} \left(G_{rrra} \!\!+\! N_1 G_{arra} \!\!+\! N_2 G_{rara} \!\!+\! N_3 G_{rraa}\!\!+\! N_1 N_2 G_{aara} \!\!+\! N_1 N_3 G_{araa}\!\!+\! N_2 N_3 G_{raaa}\right)  \nonumber\\
G_{AARA}&=&\frac{1}{C(N_1,N_2,N_4)} \left(G_{rrar} \!\!+\!  N_1 G_{arar} \!\!+\! N_2 G_{raar} \!\!+\! N_4 G_{rraa}\!\!+\! N_1 N_2 G_{aaar} \!\!+\! N_1 N_4 G_{araa}\!\!+\! N_2 N_4 G_{raaa}\right) \nonumber\\
G_{ARAA}&=&\frac{1}{C(N_1,N_3,N_4)} \left(G_{rarr} \!\!+\!  N_1 G_{aarr} \!\!+\! N_3 G_{raar} \!\!+\! N_4 G_{rara}\!\!+\! N_1 N_3 G_{aaar} \!\!+\! N_1 N_4 G_{aara}\!\!+\! N_3 N_4 G_{raaa}\right)  \nonumber\\
G_{RAAA}&=&\frac{1}{C(N_2,N_3,N_4)} \left(G_{arrr} \!\!+\!  N_2 G_{aarr} \!\!+\! N_3 G_{arar} \!\!+\! N_4 G_{arra}\!\!+\! N_2 N_3 G_{aaar} \!\!+\! N_2 N_4 G_{aara}\!\!+\! N_3 N_4 G_{araa}\right)  \nonumber\\
G_{AAAA}&=&\frac{1}{8\;\; n(-k_1)\; n(-k_2) \; n(-k_3)\; n(-k_4)}  \left( G_{rrrr} + N_1 G_{arrr} + N_2 G_{rarr} + N_3 G_{rrar} + N_4 G_{rrra} \right. \nonumber\\
&& + N_1 N_2 G_{aarr} + N_1 N_3 G_{arar} + N_1 N_4 G_{arra} 
~~ + N_2 N_3 G_{raar} + N_2 N_4 G_{rara} + N_3 N_4 G_{rraa}\nonumber\\
&&\left. + N_1 N_2 N_3 G_{aaar} + N_1 N_2 N_4 G_{aara}  + N_1 N_3 N_4 G_{araa} + N_2 N_3 N_4 G_{raaa} \right)\nonumber
\eea
Applying KMS conditions we have:
\begin{eqnarray}
G_{RRRA}&=&\frac{1}{4} \left(G^*_{rrra} \!\!+\! N_1 G^*_{arra} \!\!+\! N_2 G^*_{rara} \!\!+\! N_3 G^*_{rraa}\!\!+\! N_1 N_2 G^*_{aara} \!\!+\! N_1 N_3 G^*_{araa}\!\!+\! N_2 N_3 G^*_{raaa}\right) 
\label{RAintermsofKeldyshtwo4pttwo}\\
G_{RRAR}&=&\frac{1}{4} \left(G^*_{rrar} \!\!+\!  N_1 G^*_{arar} \!\!+\! N_2 G^*_{raar} \!\!+\! N_4 G^*_{rraa}\!\!+\! N_1 N_2 G^*_{aaar} \!\!+\! N_1 N_4 G^*_{araa}\!\!+\! N_2 N_4 G^*_{raaa}\right) 
\nonumber \\
G_{RARR}&=&\frac{1}{4} \left(G^*_{rarr} \!\!+\!  N_1 G^*_{aarr} \!\!+\! N_3 G^*_{raar} \!\!+\! N_4 G^*_{rara}\!\!+\! N_1 N_3 G^*_{aaar} \!\!+\! N_1 N_4 G^*_{aara}\!\!+\! N_3 N_4 G^*_{raaa}\right)  
\nonumber \\
G_{ARRR}&=&\frac{1}{4} \left(G^*_{arrr} \!\!+\!  N_2 G^*_{aarr} \!\!+\! N_3 G^*_{arar} \!\!+\! N_4 G^*_{arra}\!\!+\! N_2 N_3 G^*_{aaar} \!\!+\! N_2 N_4 G^*_{aara}\!\!+\! N_3 N_4 G^*_{araa}\right) 
\nonumber \\
G_{RRAA}&=&-\frac{1}{2}
    \left( G^*_{rraa} \!\!+\!\! N_1 G^*_{araa} \!\!+\!\! N_2 G^*_{raaa} \right) \nonumber\\
G_{RARA}&=&-\frac{1}{2}
    \left(  G^*_{rara} \!\!+\!\! N_1 G^*_{aara} \!\!+\!\! N_3 G^*_{raaa} \right) \nonumber\\
G_{RAAR}&=&-\frac{1}{2}
    \left(G^*_{raar} \!\!+\!\! N_1 G^*_{aaar} \!\!+\!\! N_4 G^*_{raaa}  \right) \nonumber\\
G_{ARRA}&=&-\frac{1}{2}
     \left( G^*_{arra} \!\!+\!\! N_2 G^*_{aara} \!\!+\!\! N_3 G^*_{araa} \right) \nonumber\\
G_{ARAR}&=&-\frac{1}{2}
     \left(G^*_{arar} \!\!+\!\! N_2 G^*_{aaar} \!\!+\!\! N_4 G^*_{araa}  \right) \nonumber\\
G_{AARR}&=&-\frac{1}{2}
      \left(G^*_{aarr} \!\!+\!\! N_3 G^*_{aaar} \!\!+\!\! N_4 G^*_{aara}  \right)\nonumber\\
G_{AAAR}&=&G^*_{aaar} \nonumber\\
G_{AARA}&=&G^*_{aara} \nonumber\\
G_{ARAA}&=&G^*_{araa} \nonumber\\
G_{RAAA}&=&G^*_{raaa} \nonumber\\
G_{AAAA}&=&0 \nonumber
\eea

\section{KMS Conditions} \label{KMS}
The KMS conditions are a set of relations between the various
components of a given $n$-point function that hold at equilibrium. The KMS conditions are
often useful for simplifying the expressions that result from the
contractions over indices, in any representation of real time statistical field theory.
The KMS conditions have the simplest structure when expressed in the
R/A basis.

\subsection{Keldysh Basis}

The KMS conditions in the Keldysh representation can be written:
\begin{eqnarray}
\label{general_KMS_physical}
&&C_{n_a}(\{N_i \delta_{\alpha^\prime_i,a}\})
\sum_{\alpha_l\in\{r,a\}} \left( \prod_{l=1}^{n}  
	\left[ \tilde N(k_l;\alpha_{l})\delta_{\alpha^\prime_l\,r}
		+ \delta_{\alpha^\prime_l\,a}\delta_{\alpha_l a}\right]\right)  
	G_{ \alpha_{1} \cdots \alpha_n}  \\
= 
&&C_{n_r}(\{N_i \delta_{\alpha^\prime_i,r}\})
\sum_{\alpha_l\in\{r,a\}} \left( \prod_{l={1}}^{n}  
	\left[ \tilde N(k_l;\alpha_{l})\delta_{\alpha^\prime_l\,a}
		+ \delta_{\alpha^\prime_l\,r}\delta_{\alpha_l a}\right]\right) 
	G^*_{\alpha_{1} \cdots \alpha_n }\nonumber
\end{eqnarray}
In this master equation, the set of variables $\{\alpha^\prime_i\}$ are
external variables that are not summed over. Thus, in principle,
(\ref{general_KMS_physical}) contains $2^n$ equations which come from
the $2^n$ choices of the sets $\{\alpha^\prime_i\}$. In fact, one half
of these equations is the complex conjugate of the other half and thus
we have $2^{n-1}$ KMS conditions for each $n$-point function.  We give
several examples below:\\

\xx [$n=2$]

\xx For the 2-point function we obtain $2^{n-1}\Big|_{n=2} = 2$ equations:
\bea
\label{kms-2-K}
 D_{ra}(p) = D^*_{ar}(p)~~&{\rm or}&~~D_{ret}(p) = D_{adv}^*(p)\\
D_{rr}(p) = N(p)(D_{ra}(p)-D_{ar}(p))~~&{\rm or}&~~ D_{sym}(p) = N(p)(D_{ret}(p)-D_{adv}(p))\,.\nonumber
\eea

\xx [$n=3$]

\xx There are $2^{n-1}\Big|_{n=3}= 4$ independent KMS conditions which are:
\begin{eqnarray}
\label{kms-3-K}
&&G_{rrr}+ N_1 G_{arr} + N_2 G_{rar} + N_3 G_{rra} + N_1 N_2 G_{aar} + N_1 N_3 G_{ara} + N_2 N_3 G_{raa}= 0\\
&&G_{rra} + N_1 G_{ara} + N_2 G_{raa} = (N_1 + N_2) G^*_{aar} \nonumber\\
&&G_{rar} + N_1 G_{aar} + N_3 G_{raa} = (N_1 + N_3) G^*_{ara} \nonumber\\
&&G_{arr} + N_2 G_{aar} + N_3 G_{ara} = (N_2 + N_3) G^*_{raa} \nonumber
\end{eqnarray}

\xx [$n=4$]

\xx There are $2^{n-1}\Big|_{n=4}= 8$ independent KMS conditions which are:
\begin{eqnarray}
&&G_{rrrr} + N_1 G_{arrr} + N_2 G_{rarr} + N_3 G_{rrar} + N_4 G_{rrra}  + N_1 N_2 G_{aarr} + N_1 N_3 G_{arar} + N_1 N_4 G_{arra} \\
&&~~ + N_2 N_3 G_{raar} + N_2 N_4 G_{rara} + N_3 N_4 G_{rraa} + N_1 N_2 N_3 G_{aaar} + N_1 N_2 N_4 G_{aara} \nonumber\\
&&~~ + N_1 N_3 N_4 G_{araa} + N_2 N_3 N_4 G_{raaa}=0
\label{kms-4-K}\nonumber\\[5mm]
&&G_{rrra} \!\!+\! N_1 G_{arra} \!\!+\! N_2 G_{rara} \!\!+\! N_3 G_{rraa}\!\!+\! N_1 N_2 G_{aara} \!\!+\! N_1 N_3 G_{araa}\!\!+\! N_2 N_3 G_{raaa} \nonumber\\
&&~~ =(1 + N_1 N_2 + N_1 N_3 + N_2 N_3) G^*_{aaar} \nonumber\\[2mm]
&&G_{rrar} \!\!+\!  N_1 G_{arar} \!\!+\! N_2 G_{raar} \!\!+\! N_4 G_{rraa}\!\!+\! N_1 N_2 G_{aaar} \!\!+\! N_1 N_4 G_{araa}\!\!+\! N_2 N_4 G_{raaa} \nonumber\\
&&~~=(1 + N_1 N_2 + N_1 N_4 + N_2 N_4) G^*_{aara} \nonumber\\[2mm]
&& G_{rarr} \!\!+\!  N_1 G_{aarr} \!\!+\! N_3 G_{raar} \!\!+\! N_4 G_{rara}\!\!+\! N_1 N_3 G_{aaar} \!\!+\! N_1 N_4 G_{aara}\!\!+\! N_3 N_4 G_{raaa} \nonumber\\
&&~~=(1 + N_1 N_3 + N_1 N_4 + N_3 N_4) G^*_{araa} \nonumber\\[2mm]
&& G_{arrr} \!\!+\!  N_2 G_{aarr} \!\!+\! N_3 G_{arar} \!\!+\! N_4 G_{arra}\!\!+\! N_2 N_3 G_{aaar} \!\!+\! N_2 N_4 G_{aara}\!\!+\! N_3 N_4 G_{araa} \nonumber\\
&&~~=(1 + N_2 N_3 + N_2 N_4 + N_3 N_4) G^*_{raaa}\nonumber\\[5mm]
&&(N_3 \!\!+\!\! N_4) \left( G_{rraa} \!\!+\!\! N_1 G_{araa} \!\!+\!\! N_2 G_{raaa} \right)\! = \! (N_1\!\!+\!\!N_2)\left(G^*_{aarr} \!\!+\!\! N_3 G^*_{aaar} \!\!+\!\! N_4 G^*_{aara}  \right) \nonumber\\[2mm]
&&(N_2 \!\!+\!\! N_4) \left( G_{rara} \!\!+\!\! N_1 G_{aara} \!\!+\!\! N_3 G_{raaa} \right)\! = \! (N_1\!\!+\!\!N_3)\left(G^*_{arar} \!\!+\!\! N_2 G^*_{aaar} \!\!+\!\! N_4 G^*_{araa}  \right) \nonumber\\[2mm]
&&(N_1 \!\!+\!\! N_4) \left( G_{arra} \!\!+\!\! N_2 G_{aara} \!\!+\!\! N_3 G_{araa} \right)\! = \! (N_2\!\!+\!\!N_3)\left(G^*_{raar} \!\!+\!\! N_1 G^*_{aaar} \!\!+\!\! N_4 G^*_{raaa}  \right)\nonumber
\end{eqnarray}

\xx [$n=5$]

\xx There are $2^{n-1}\Big|_{n=5}= 16$ independent KMS conditions. We write down three representative ones:
\bea
&&G_{rrrrr} + N_1 G_{arrrr} + N_2 G_{rarrr} + N_3 G_{rrarr} + N_4 G_{rrrar} + N_5 G_{rrrra} \\
&&~~ +N_1 N_2 G_{aarrr} + N_1 N_3 G_{ararr} + N_1 N_4 G_{arrar} + N_1 N_5 G_{arrra} + N_2 N_3 G_{raarr} \nonumber\\
&&~~+  N_2 N_4 G_{rarar} + N_2 N_5  G_{rarra} + N_3 N_4 G_{rraar} + N_3 N_5 G_{rrara} + N_4 N_5 G_{rrraa} \nonumber\\
&&~~+  N_1 N_2 N_3 G_{aaarr} + N_1 N_2 N_4 G_{aarar}+ N_1 N_2 N_5  G_{aarra} + N_1 N_3 N_4 G_{araar} \nonumber\\ 
&&~~+N_1 N_3 N_5 G_{arara} + N_1 N_4 N_5 G_{arraa} + N_2 N_3 N_4 G_{raaar} + N_2 N_3 N_5 G_{raara} \nonumber \\ 
&&~~+N_2 N_4 N_5 G_{raraa} + N_3 N_4 N_5  G_{rraaa} + N_1 N_2 N_3 N_4 G_{aaaar} + N_1 N_2 N_3 N_5 G_{aaara}\nonumber \\ 
&&~~+N_1 N_2 N_4 N_5 G_{aaraa} +  N_1 N_3 N_4 N_5 G_{araaa}+ N_2 N_3 N_4 N_5 G_{raaaa}=0  \nonumber \\[2mm]
&& G_{rrrra} + N_1 G_{arrra} + N_2 G_{rarra} + N_3 G_{rrara} + N_4 G_{rrraa} + N_1 N_2 G_{aarra}\nonumber \\ 
&&~~+N_1 N_3 G_{arara} + N_1 N_4 G_{arraa}  + N_2 N_3 G_{raara} +  N_2 N_4 G_{raraa} + N_3 N_4 G_{rraaa}\nonumber \\
&&~~+  N_1 N_2 N_3 G_{aaara} + N_1 N_2 N_4 G_{aaraa}+ N_1 N_3 N_4 G_{araaa}+ N_2 N_3 N_4 G_{raaaa} \nonumber\\
&&~~= 
(N_1 + N_2 + N_3 + N_4 + N_1 N_2 N_3 + N_1 N_2 N_4 + N_1 N_3 N_4 + N_2 N_3 N_4) G^*_{aaaar}\nonumber\\[2mm]
&&(N_4 + N_5) \left(G_{rrraa} + N_1 G_{arraa} + N_2 G_{raraa} + N_3 G_{rraaa} + N_1 N_2 G_{aaraa}+ N_1 N_3 G_{araaa} + N_2 N_3 G_{raaaa}\right) \nonumber\\
&&~~=
(1+N_1 N_2 + N_1 N_3 + N_2 N_3 )\left( G^*_{aaarr} + N_4 G^*_{aaaar} + N_5 G^*_{aaara} \right) \nonumber
\eea

The KMS conditions for the vertex
functions have almost exactly the same form.  The master equation is obtained from 
(\ref{general_KMS_physical}) by interchanging the indices $r$ and $a$:
\begin{eqnarray}
\label{general_KMS_physical-2}
&&C_{n_r}(\{N_i \delta_{\alpha^\prime_i,r}\})
\sum_{\alpha_l\in\{r,a\}} \left( \prod_{l=1}^{n}  \left[ \hat N(k_l;\alpha_{l})\delta_{\alpha^\prime_l\,a}+ \delta_{\alpha^\prime_l\,r}\delta_{\alpha_l r}\right]\right)  \Gamma^{ \alpha_{1} \cdots \alpha_n} \\
= 
&&C_{n_a}(\{N_i \delta_{\alpha^\prime_i,a}\})
\sum_{\alpha_l\in\{r,a\}} \left( \prod_{l={1}}^{n}  \left[ \hat N(k_l;\alpha_{l})\delta_{\alpha^\prime_l\,r}+ \delta_{\alpha^\prime_l\,a}\delta_{\alpha_l r}\right]\right) (\Gamma^*)^{\alpha_{1} \cdots \alpha_n }\nonumber 
\end{eqnarray}

\subsection{R/A Basis}

The KMS conditions have an even simpler form in the R/A
representation. This can be anticipated by comparing the general
structure of (\ref{RAonenewest}) and (\ref{general_KMS_physical}). The general expression for the KMS conditions in the
R/A basis is:
\begin{equation}
2^{n_R-n_A}\,C_{n_A}(\{N_i\delta_{X_i,A}\})\; G_{X_1\cdots X_n } 
 = (-1)^{n} C_{n_{\bar A}}(\{N_i\delta_{X_i,\bar A}\})\;  G^*_{   \bar{X}_{1} \cdots \bar{X}_n  }  \label{general_KMS_RA}
\end{equation}
This master equation contains a series of equations where the variables $X_i$ take all possible combinations of the values $\{R,A\}$. We use the notation: $\bar{A}=R$, $\bar{R}=A$ so that ${n}_{\bar R}={n}_{A}$ and ${n}_{\bar A}={n}_R$. We give several examples below:\\

\xx [$n=2$] For the 2-point functions we have: 
\bea
\label{kms-2-RA}
D_{AR}(p) = D^*_{RA}(p)\,;~~D_{AA} = 0.
\eea

\xx [$n=3$] For the 3-point functions we have:
\begin{eqnarray}
G_{AAA} = 0~~~~~~~~~~~~~~~~~~~~~~~~~\;&& \\
2C(N_3)G_{RRA}= -\, C(N_1,N_2)\; G^*_{AAR} &~\to ~& 2 G_{RRA}= -\, (N_1+N_2)\; G^*_{AAR}\nonumber \\
2 C(N_2)G_{RAR}= -\, C(N_1,N_3)\; G^*_{ARA} &~\to ~& 2 G_{RAR}= -\, (N_1+N_3)\; G^*_{ARA}\nonumber \\
2 C(N_1)G_{ARR}= -\,C(N_2,N_3)\; G^*_{RAA} &~\to~& 2 G_{ARR}= -\,(N_2+N_3)\; G^*_{RAA}\nonumber
\end{eqnarray}
where the arrows indicate the results that are obtained by replacing the $C$-functions by their definitions in terms of thermal functions (see Eqn. (\ref{C_definition})).\\

\xx [$n=4$] For the 4-point functions we have:

\begin{eqnarray}
G_{AAAA}=0~~~~~~~~~~~~~~~~~~~~~~~ &&\\
C(N_3,N_4)G_{RRAA}=  C(N_1,N_2)\; G^*_{AARR} &~\to ~& (N_3+N_4)G_{RRAA}=  (N_1+N_2)\; G^*_{AARR}\nonumber \\
C(N_2,N_4)G_{RARA}=  C(N_1,N_3)\; G^*_{ARAR} &~\to ~& (N_2+N_4)G_{RARA}=  (N_1+N_3)\; G^*_{ARAR}\nonumber \\
C(N_2,N_3)G_{RAAR}=  C(N_1,N_4)\; G^*_{ARRA} &~\to ~& (N_2+N_3)G_{RAAR}=  (N_1+N_4)\; G^*_{ARRA}\nonumber \\
4G_{ARRR}=  C(N_2,N_3,N_4)\; G^*_{RAAA} &~\to ~& 4G_{ARRR}=  (1+N_2N_3+N_3N_4+N_4N_2)\; G^*_{RAAA}\nonumber \\
4G_{RARR}=  C(N_1,N_3,N_4)\; G^*_{ARAA} &~\to ~& 4G_{RARR}=  (1+N_1N_3+N_3N_4+N_4N_1)\; G^*_{ARAA}\nonumber \\
4G_{RRAR}=  C(N_1,N_2,N_4)\; G^*_{AARA} &~\to ~& 4G_{RRAR}=  (1+N_1N_2+N_2N_4+N_4N_1)\; G^*_{AARA}\nonumber \\
4G_{RRRA}=  C(N_1,N_2,N_3)\; G^*_{AAAR} &~\to ~& 4G_{RRRA}=  (1+N_1N_2+N_2N_3+N_3N_1)\; G^*_{AAAR}\nonumber
\end{eqnarray}
\xx [$n=5$] For the 5-point functions we give a few examples:
\begin{eqnarray}
G_{AAAAA}&=&0\\
C(N_3,N_4,N_5)G_{RRAAA}&=& - 2C(N_1,N_2)\; G^*_{AARRR}\nonumber\\
~\to ~ (1+N_3N_4+N_4N_5+N_5N_3)G_{RRAAA}&=&  -2(N_1+N_2)\; G^*_{AARRR}\nonumber \\
8G_{RRRRA}&=& - C(N_1,N_2,N_3,N_4)\; G^*_{AAAAR}\nonumber\\
~\to ~ 8G_{RRRRA} &=& - (N_1+N_2+N_3+N_4+N_1 N_2 N_3+N_2 N_3 N_4+ N_3 N_4 N_1+N_4 N_3 N_2)\; G^*_{AAAAR}\nonumber 
\end{eqnarray}

\xx Note that in every case one of the KMS equations requires the vanishing of the $G_{A\cdots A}$ component of the $n$-point function.  The rest of the KMS equations are  simple
relations between pairs of off-diagonal components.\\

\xx The KMS conditions for the vertex functions have almost 
exactly the same form.  They are obtained from the same master equation, with the indices $R$ 
and $A$ interchanged:
\begin{equation}
2^{n_A-n_R}\,C_{n_R}(\{N_i\delta_{X_i,R}\})\;
\Gamma^{X_1\cdots X_n } = (-1)^{n}
C_{n_{\bar R}}(\{N_i\delta_{X_i,\bar R}\})\; (\Gamma^*)^{ \bar{X}_{1}
\cdots \bar{X}_n } \label{general_KMS_RA-2}
\end{equation}

\section{A program to perform the contraction of indices}
\label{program}

\subsection{Description of the Program}

Our program calculates the integrand for any diagram by contracting
indices. The calculation can be done in the 1-2, Keldysh or R/A basis.
The basic strategy of the program is to treat a Feynman diagram as a
tensor product of propagators, and bare and corrected vertices.  In
principle, this type of calculation can be done by hand, but when
working with more than a few indices the process becomes extremely
tedious. The symbolic manipulation program {\it Mathematica} is
ideally suited to perform this kind of tensor calculation.

The program is divided into six main sections. Only the {\bf Input}
section needs to be edited by the user. The user enters some variables
in this section, in order to specify the diagram that he wants to
calculate. The rest of the program can be immediately executed, and
the result is output in the last section. The user has the option to
output the results to a file. The results are also defined
functionally. Some basic functions are defined within the program that
can be used interactively to manipulate the result.

When working in the RA basis the program always assumes thermal
distribution functions, and thus the result is only valid in
equilibrium. The user has the option to implement the KMS conditions,
since both forms of the result can be useful. In the Keldysh basis,
thermal distribution forms are only assumed if the KMS conditions are
used. In the 1-2 basis the KMS conditions cannot be implemented in the
current version of the program. Note that we have not used different
notation to indicate equilibrium and non-equilibrium distributions
(see Eqns. (\ref{equib1}) and (\ref{therm-basic})): the output of the
program will always contain distribution functions written in the form
$n_p$ or $N_P$.

The program is designed to contract CTP indices and is written for scalar bosons. We set all coupling constants to one and use the notation:
\par\begin{figure}[H]
\begin{center}
\includegraphics[width=2cm]{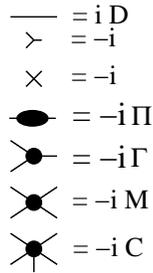}
\end{center}
\caption{Definitions of notation for propagator and vertices}
 \label{idefn}
\end{figure}
\xx Note that additional numerical factors are introduced in the Keldysh representation, as defined in (\ref{Keldyn}). 
The program can be used for other field theories with appropriate adjustments. The appropriate coupling constant(s) must be inserted by the user.   In addition, any dirac, lorentz, or other group structure must be separately handled by the user. \\

We describe briefly the main sections of the program:

\begin{enumerate}
\item The {\bf Initialization} section inputs the necessary {\it Mathematica} package.

\item  The {\bf Input} section is edited by the user to input specific parameters corresponding to the diagram he wants to calculate. This process is described in detail below.

\item The {\bf Definitions} section establishes some basic definitions that will be used throughout the program. 

\item The {\bf Find Loops} section identifies all closed loops in the diagram. The variable ({\it loopzerosubs}) is printed out. The usefulness of this variable is explained in point 11 below. 

\item The {\bf Calculate Diagram} section performs the calculation. 

\item The {\bf Results} section outputs the results of the calculations and performs some basic manipulations to simplify them. 
\end{enumerate}

\subsection{{\bf Input} section}

We illustrate the {\bf Input} section of the program with an example. Consider the diagram in Fig. \ref{leaf-A}.
\par\begin{figure}[H]
\begin{center}
\includegraphics[width=4cm]{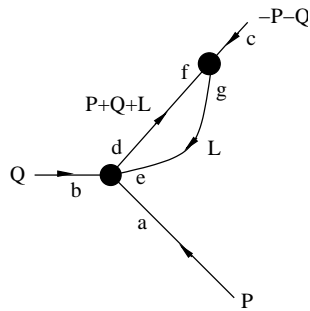}
\end{center}
\caption{Example Feynman diagram.  Uppercase letters are momenta and lowercase letters
label indices of the propagators and the vertices.}
 \label{leaf-A}
\end{figure}

\xx The {\bf Input} section for this diagram is reproduced
on the page that follows the itemized list below. Each entry is explained in the order it appears in the {\bf Input} section.

\begin{enumerate}
\item {\tt number of external legs, external momenta, external indices}: Specify the number of external legs, and the momentum and corresponding index for each leg.  
\item {\tt number of internal indices, internal indices}: Specify the number of internal indices and list them.  

\item {\tt list of momenta, list of indices}: For each bare vertex, list the incoming momentum for each leg and the corresponding index. 

\item  {\tt list of momenta, list of indices}: 
For each corrected 3-point vertex, corrected 4-point vertex and corrected 5-point vertex, list the incoming momentum for each leg and the corresponding index. 

\item {\tt list of momenta, list of indices}: List the momentum and the corresponding pair of initial and 
final indices for each propagator.  

$\bullet$ In this example, there is one corrected 3-point vertex and one corrected 4-point vertex. The corresponding momentum arguments are listed in double set brackets, indicating that the one vertex is the first in a list of length one. The same notation is used in any case where multiple entries would create nested lists. This includes the momenta and indices of bare or corrected $n$-point vertices, with $n\ge 3$, and the indices of propagators. The momenta for any number of propagators (even if there is only one) are listed in single set brackets. 

\item {\tt choose basis (onetwo, Keldysh or RA)}: Indicate the basis  as
either {\it onetwo}, {\it Keldysh} or {\it RA}.  

\item {\tt combination(s) to be evaluated: 
    ie. {ra, ...} or {AR, .} or All}: Indicate the set of external indices of the diagrams that are to be calculated or simply specify {\it All}.  

\item {\tt Simplify result? (option to use the Mathematica function `Simplify')}: The user can choose to have the program apply the {\it Mathematica} function {\it Simplify} to the result ({\it simplifyIt = yes}). For smaller diagrams this produces neater results. For larger diagrams it can lead to significantly longer running times, without producing a result that is much more compact.

\item {\tt  use the KMS conditions}: In the Keldysh or RA bases, one can choose to enforce the KMS conditions ({\it useKMSconditions=yes}). In the 1-2 basis choosing ({\it useKMSconditions=yes}) has no effect on the output. If the user has chosen the Keldysh basis, he can enforce the KMS conditions for the propagator, but not for higher $n$-point vertices ({\it removeFs=yes}).  Note that if the user chooses {\it useKMSconditions=yes} and {\it removeFs=no}, the first choice over-rides the second, and KMS conditions are used for all $n$-point functions including propagators. 

\item {\tt express answer in terms of RA expressions}: In the Keldysh basis, if the calculation was done using the KMS conditions, the user
can request that the right-hand side of the final expression be written in the R/A basis ({\it InTermsOfRA = yes}), since this basis frequently produces more compact results. The default value of this parameter is ``no'' and produces results in which both sides of the equation are written in the same basis.

\item {\tt remove terms which are zero after integration}: One can choose to eliminate terms 
which will vanish after integration 
({\it removeZeros=yes}). This option only works in the Keldysh or RA bases (in the 1-2 basis choosing 
({\it removeZeros=yes}) has no effect on the output). The program identifies closed loops in the diagram, and then looks for terms in which all of the propagators that form a closed loop have poles on the same side of the real axis. Choosing  ({\it removeZeros=yes}) will cause the program to set all of these terms (except for tadpoles) to zero. The program automatically exempts tadpole loops, so that they are not incorrectly set to zero. The list of combinations of propagators which will be removed by this option are stored in ({\it loopzerosubs}) and printed in the {\bf Find Loops} section of the program. The propagators that correspond to tadpoles are also listed. One can compare these lists to the original diagram as a check of the information given in the {\bf Input} section. For our example we have: 
\begin{eqnarray}
G_{\text{ra}}(L) G_{\text{ra}}(L+P+Q)\to 0\,;~~ G_{\text{ar}}(L)
   G_{\text{ar}}(L+P+Q)\to 0
\end{eqnarray}
The two expressions correspond to the one loop in our example diagram, with the momentum routed clockwise or counter-clockwise.

\item {\tt replace C functions in terms of N functions}: If the KMS conditions are used, the program does the calculation in terms of the $C$-functions defined in Eqn. (\ref{C_definition}). One can choose to replace these 
$C$-functions with their definitions in terms of thermal functions ({\it replaceCs=yes})  - see Eqn. (\ref{C_definition}). 
 
\item {\tt specify name of output file using quotes}: The user can specify a filename to output the results to, or specify an
empty string (``'') for no output file.

\item {\tt Directory for output file is directory of this notebook as specified below}: The directory of the output file is set to be the same as the directory of the original notebook. The user can specify another directory for the output file by rewriting this line.
\end{enumerate}

\par\begin{figure}
\begin{center}
\includegraphics[width=20cm]{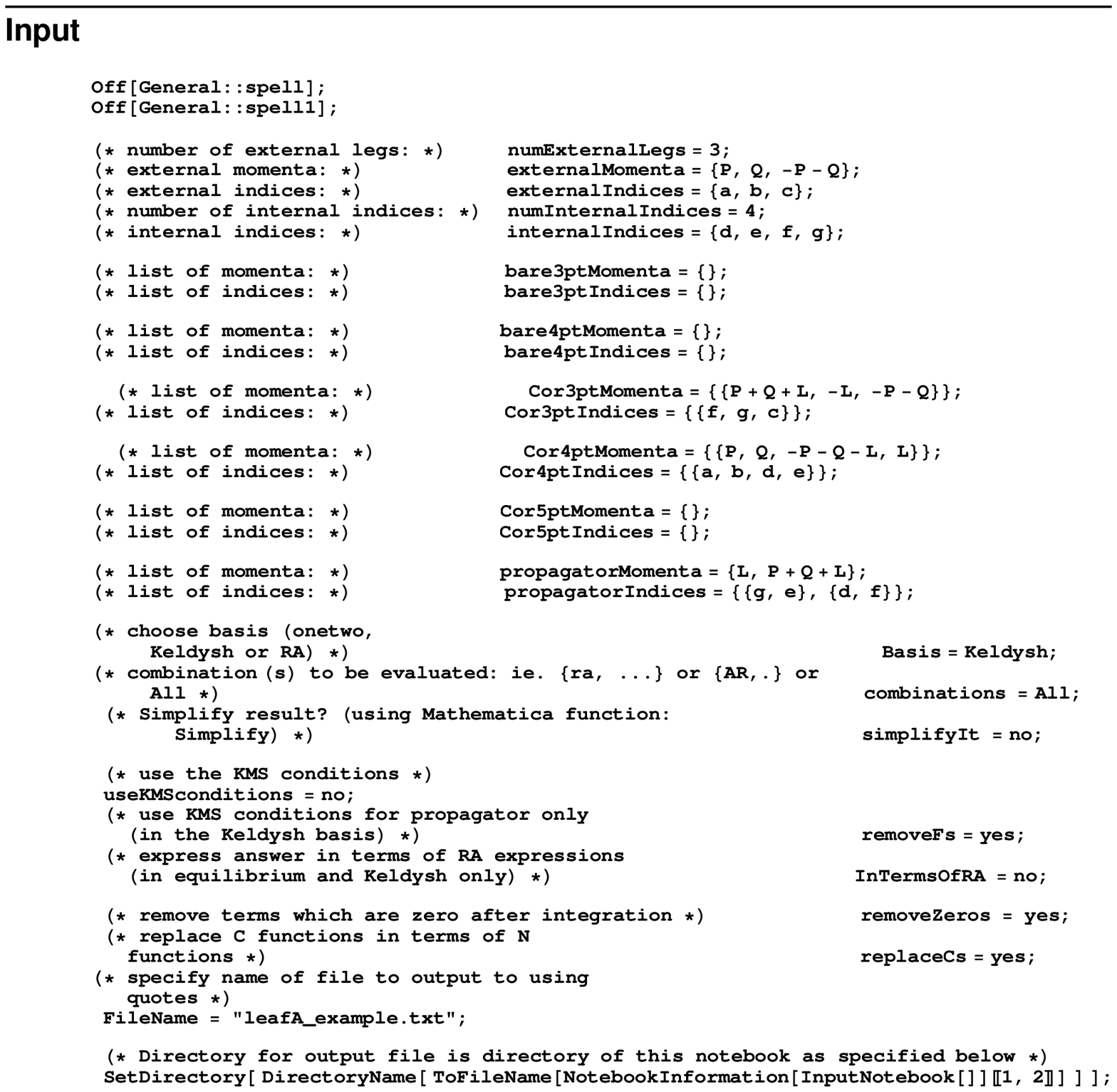}
\end{center}
\caption{Input cell for Program}
 \label{example_input}
\end{figure}
\subsection{Error messages}

Some checks have
been implemented to detect possible input mistakes, and give error
messages indicating the nature of the problem.  A few examples are given below:

\begin{enumerate}

\item If the
incoming momenta for any of the internal vertices, or for the external legs, do
not sum to zero, the error message appears: ``Momentum non-conserving vertex
detected.''  

\item If the combinations of external indices requested using ``{\tt combination(s) to be evaluated}'' do not correspond to the basis specified using ``{\tt choose basis (onetwo, Keldysh or RA)},'' an error message appears. For example, choosing the RA basis and requesting the combination $rra$ (corresponding to the vertex function $\Gamma^{rra}$) produces the error message: ``Initialization Failed: 
    Combination {\it rra} is not specified in the RA basis.''

\item If the combinations of external indices requested do not have the same number of variables as the number of external legs, an error message will appear. For example, calculating a 3-point vertex function (as in our example) and requesting the combination $rraa$ (corresponding to the 4-point vertex $\Gamma^{rraa}$) produces the error message: ``Initialization Failed: 
    Combination $rraa$ does not have the correct number of indices.''
    
\item Each external index must appear one time only as a vertex index. Each internal index must appear one time as a vertex index and one time as a propagator index. If the indices are incorrectly entered, an error message will appear.  

\item Because of the fact that vertices are defined with incoming momenta, for each propagator, the momentum corresponding to the second index
must be the same as the momentum for the vertex leg with the matching index. Similarly,
the momentum corresponding to the first index
must be the negative of the momentum for the vertex leg with the matching index.
If the momentum variables are incorrectly entered, an error message will appear.  
\end{enumerate}

\subsection{{\bf Interactive Manipulation of the Results}}

The result of the calculation is stored in the internal variable ``{\it diagram}.''  The user can extract and manipulate the results in several ways.

\begin{enumerate}

\item One can extract a specific
result using the function {\it DIAGRAM}. For our example, evaluating $DIAGRAM[a,r,r]$ 
gives the result for $\Gamma^{arr}$ in the Keldysh representation:
\bea
\label{ex1}
&&-\frac{1}{2} i G_{\text{ar}}(L) G_{\text{ra}}(L+P+Q) \Gamma ^{\text{arrr}}(P,Q,L,-L-P-Q) \\
&&~~~~\Big[\Gamma ^{\text{aar}}(L+P+Q,-L,-P-Q)-N_L \Gamma
   ^{\text{arr}}(L+P+Q,-L,-P-Q)\nonumber\\
&&~~~~~~+N_{L+P+Q} \Gamma ^{\text{rar}}(L+P+Q,-L,-P-Q)\Big]\nonumber
\eea
(the prefactor $\int \frac{d^4l}{(2\pi)^4}$ is not explicitly written).

\item In Eqn. (\ref{ralist}) we define notation for the Keldysh basis so that each string of indices of the form $rrar\cdots$ corresponds to a single numerical index. A function called GAMMA is defined that will automatically supply the result of the calculation for a given numerical index. 
All vertices are  translated into the notation defined in Eqn. (\ref{ralist}). Propagators are written using the notation $D_{ra}(P)=D_{ret}(P)  := r(P)$; $D_{ar}(P)=D_{adv}(P)  := a(P)$. The
function GAMMA is illustrated below. From Eqn. (\ref{ralist}), the index `2' corresponds to the string `arr'  for a 3-point vertex function. 
For our example, evaluating $\mbox{GAMMA}[2]$ produces the translation of the result for $\Gamma^{arr}$ that is given in (\ref{ex1}):
\bea
\label{ex2}
&&-\frac{1}{2} i a(L)r(L+S) \text{M}(2,P,Q,L,-L-S)  \\
&&~~~~ \Big[\Gamma (4,L+S,-L,-S)-N_L \Gamma (2,L+S,-L,-S)+N_{L+S} \Gamma (3,L+S,-L,-S)\Big]\nonumber
\eea

\end{enumerate}

\section{The QED Ward identity for the 4-point function}
\label{WI}

In this section, we present a calculation of the ward identity for the
QED 3-point vertex function, and the QED 4-point vertex function
involving two photons and two fermions.  Throughout this section we
work in the Keldysh representation. 
As explained in the beginning of section \ref{program}, the goal of our program is provide compact results for the expressions that result
from summing over CTP indices. The dirac structure of fermions and any group structure associated with all fields, must be handled separately
by the user.  For the calculations done in this paper, there are no additional complications. The ward identities are derived by comparing
groups of integrands, without evaluating the integrals themselves. Consequently, we can simply suppress all dirac and lorentz indices.
For example, the result for the contribution to $\Sigma(2,p)$ shown in Fig. \ref{Sigma} is given by (\ref{PIres}). The first term  is 
$-\frac{i }{2}\int \frac{d^4l}{(2\pi)^4}a(l)r(l+p)\Gamma(2,p,l+p)\Gamma(4,l+p,p)$. Including all indices, this factor gives a contribution:
\bea
 -\frac{i }{2}\int \frac{d^4l}{(2\pi)^4} a_{\mu\nu}(l) \Gamma^\mu_{\beta \beta'}(2,l,l+p)r_{\beta' \alpha'}(l+p)\Gamma^\nu_{\alpha'\alpha}(4,l+p,p)
\eea
where $\{\mu,\nu\}$ are Lorentz indices and $\{\alpha,\alpha',\beta,\beta'\}$ are Dirac indices. Throughout the rest of this section we will
leave all integrands in the form described above, with dirac and lorentz indices suppressed. In addition, we define $k + p + q = u$, $k + p = t$, $p + q = s$ and use the shorthand notation: $D_{ret}(p) = r(p)$, $D_{adv}(p) = a(p)$, $D_{sym}(p) = f(p)$.

In order to simplify the notation for the vertices, we replace each combination of the indices $\{r,a\}$ by a single numerical index:
\bea
\Gamma^{(n)~\alpha_1\alpha_2\cdots \alpha_n}(p_1,p_2, \cdots p_n) = \Gamma^{(n)}(i,p_1,p_2, \cdots p_n)
\eea 
We assign 
the choices of the variables $\alpha_1 \alpha_2 \cdots \alpha_n$ to the variable $i$ using the vector
\bea 
\label{ralist}
V_n = 
\Big(
\begin{array}{c}
r_n \\
a_n
\end{array}
\Big) \cdots \otimes
\Big(
\begin{array}{c}
r_2 \\
a_2
\end{array}
\Big)\otimes
\Big(
\begin{array}{c}
r_1 \\
a_1
\end{array}
\Big)
\eea
where the symbol $\otimes$ indicates the outer product. For each $n$,
the $i$th component of the vector corresponds to a list of variables
that is assigned the number $i$.  To simplify the notation we drop the
subscripts and write a list like $r_1 r_2 a_3$ as $rra$. For clarity,
the results are listed below. \\

\xx [a] 2-point functions: $rr \ra 1$, $ar\ra 2$, $ra\ra 3$, $aa\ra 4$\\

\xx [b] 3-point functions: $rrr\ra 1$, $arr\ra 2$, $rar\ra 3$, $aar\ra 4$, $rra\ra 5$, $ara\ra 6$, $raa\ra 7$, $aaa\ra 
8$\\

\xx [c] 4-point functions: $rrrr \ra 1$, $arrr \ra 2$,  $ rarr \ra 3$,  $ aarr \ra 4$,  $ rrar \ra 5$,  $ arar \ra 6$,  $ raar \ra 7$,  $aaar \ra 8$,  $ rrra \ra 9$,  $ arra \ra 10$,  $ rara \ra 11$,  $ aara \ra 12$,  $ rraa \ra 13$,  $araa \ra 
14$,  $ raaa \ra 15$, $ aaaa \ra 16$\\

\xx [d] 5-point functions: $rrrrr \ra 1$, $arrrr \ra 2$, $rarrr \ra 3$, $aarrr \ra 4$, $rrarr \ra 5$, $ararr \ra 6$, $raarr \ra 7$, $aaarr \ra 8$, $rrrar \ra 9$, $arrar \ra 10$, $rarar \ra 11$, $aarar \ra 12$, $rraar \ra 13$, $araar \ra 14$, $raaar \ra 15$, $aaaar \ra 16$, $rrrra \ra 17$, $arrra \ra 18$, $rarra \ra 19$, $aarra \ra 20$, $rrara \ra 21$, $arara \ra 22$, $raara \ra 23$, $aaara \ra 24$, $rrraa \ra 25$, $arraa \ra 26$, $raraa \ra 27$, $aaraa \ra 28$, $rraaa \ra 29$, $araaa \ra 30$, $raaaa \ra 31$, $aaaaa \ra 32$\\

\xx We note that $i=1$ corresponds to a vertex that is identically zero, for any number of external legs, as a consequence of the general relation $\Gamma^{rrr\cdots r}=0$ which can be obtained from (\ref{general_KMS_physical-2}). \\

\noindent To reduce the number of indices we will introduce separate names for the first five vertex functions and write:
\bea
\Gamma^{(2)} = \Sigma\;;~~\Gamma^{(3)} = \Gamma\;;~~\Gamma^{(4)} = M\;;~~\Gamma^{(5)} = C \,.
\eea
The notation for the momentum arguments is as follows:
\bea
\Sigma(p_{in})\;;~~\Gamma^\mu(p_{in},p_{out})\;;~~M^{\mu\nu}(p_{in},q^\mu_1,q^\nu_2,p_{out})\;;~~C^{\mu\nu\tau}(p_{in},q^\mu_1,q^\nu_2,q^\tau_3,p_{out}) \,,
\eea
where $p_{in}$ is the momentum of the incoming fermion, $\{q_1,q_2,q_3\}$ are the momenta of the incoming photons, and $p_{out}$ is the momentum of the outgoing fermion. Note that the momentum of the photon is not written for the 3-point vertex since it can be inferred from the momenta of the fermions. Similarly, the self energy is written $\Sigma(p_{in})$ instead of $\Sigma(p_{in},-p_{in})$.

We begin by calculating the
ward identity at the bare 1-loop level. In the next section, we verify
that the same ward identities are satisfied by the complete set of
graphs involving full corrected vertices.

\subsection{Bare 1-loop Diagrams}

We start by looking at 1-loop diagrams with bare vertices, in order to determine the form of the ward identities. For the 2- and 3-point functions the graphs are shown in Figs. (\ref{bare}a) and (\ref{bare}b). For the 4-point function we have the box graph  (Fig. (\ref{bare}c)), and the crossed version of the box graph where the two external photons are interchanged. For the five point function, the basic graph is show in Fig. (\ref{bare}d). There are six versions of this graph which correspond to the six possible permutations of the three external photons.  
\par\begin{figure}[H]

\begin{center}
\includegraphics[width=10cm]{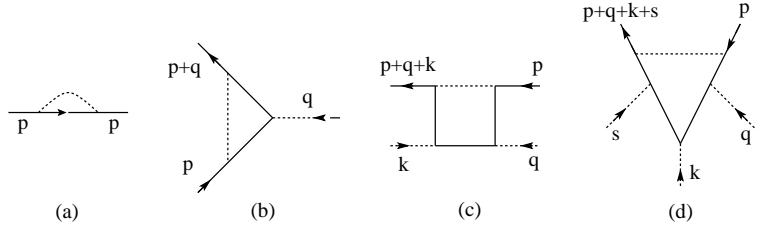}
\end{center}
\caption{One loop diagrams}
 \label{bare}
\end{figure}
We use the notation defined in Fig. \ref{idefn} and insert the appropriate coupling constants for QED. We include the numerical factor as defined in (\ref{Keldyn}):
\bea
&& -i\Sigma^{(1)} = {\rm self-energy~graph} \\
&& -i e\Gamma^{(1)} = \sqrt{2}\,\Big({\rm triangle~graph}\Big) \nonumber\\
&& -i e^2 M^{(1)} = 2\,\Big({\rm box~graph} + {\rm crossed~box~graph}\Big)\nonumber \\
&& -i e^3 C^{(1)} = 2\,\sqrt{2}\,\Big({\rm six~permutations~of~5-point~graph}\Big) \nonumber
\eea
where the superscripts refer to the loop order of the graph.
The ward identities are of the form
\bea
\label{WIbasic}
&& Q^\mu \Gamma^{(1)}_\mu (i,p,p+q) = \Sigma^{(1)} (j_1,p)-\Sigma^{(1)}(j_2,p+q) \\
&& Q^\mu M^{(1)}_{\mu\nu} (i,p,q,k,p+q+k) = \Gamma^{(1)}_\nu(j_1,p,p+k)-\Gamma^{(1)}_\nu(j_2,p + q, k + p + q)\nonumber\\
&& Q^\mu C^{(1)}_{\mu\nu\tau} (i,p,q,k,s,p+q+k+s) = M^{(1)}_{\nu\tau}(j_1,p, k, s, k + p + s)-M^{(1)}_{\nu\tau}(j_2,p + q, k, s, k + p + q + s)\nonumber
\eea
where the indices $\{i,j_1,j_2\}$ refer to the choices of the variables $\alpha_1 \alpha_2 \cdots \alpha_n$ as defined in (\ref{ralist}). We list below the sets of these indices for the 3-point, 4-point and 5-point functions. 
\bea
\label{WIbare}
 \Gamma:~~&&(2,2,2),~(3,3,2),~(4,4,1),~(5,3,3),~(6,4,4),~(7,1,4),~(8,2,3)\\[2mm]
M:~~&&(2,2,2),~(3,5,2),~(4,6,1),~(5,3,3),~(6,4,4),~(7,7,4),~(8,8,3),~(9,5,5),~(10,6,6),~(11,1,6),\nonumber\\
&&(12,2,5),~(13,7,7),~(14,8,8),~(15,3,8),~(16,4,7)\nonumber\\[2mm]
C:~~&&(2,2,2),~(3,9,2),~(4,10,1),~(5,3,3),~(6,4,4),~(7,11,4),~(8,12,3),~(9,5,5),~(10,6,6),~(11,13,6),\nonumber\\
&&(12,14,5),~(13,7,7),~(14,8,8),~(15,15,8),~(16,16,7),~(17,9,9),~(18,10,10),~(19,1,10),~(20,2,9),\nonumber\\
&&(21,11,11),~(22,12,12),~(23,3,12),~(24,4,11),~(25,13,13),
~(26,14,14),~(27,5,14),~(28,6,13),~(29,15,15),\nonumber\\
&&(30,16,16),~(31,7,16),~(32,8,15)\nonumber
\eea
We give three specific examples below. The first set of numbers in the first line of (\ref{WIbare}) is (2,2,2). Using (\ref{ralist}) and (\ref{WIbasic}) the corresponding ward identity is:
\bea
\label{example1}
&&Q^\mu \Gamma^{(1)}_\mu (2,p,p+q) = \Sigma^{(1)} (2,p)-\Sigma^{(1)}(2,p+q) \\
{\rm or}~~&&Q \cdot \Gamma^{(1)}_{arr} (p,p+q) = \Sigma^{(1)}_{ar} (p)-\Sigma^{(1)}_{ar}(p+q) \nonumber
\eea
The first set of numbers in the second line of (\ref{WIbare}) is (2,2,2).
The corresponding ward identity is
\bea
\label{example2}
&& Q^\mu M^{(1)}_{\mu\nu\tau} (2,p,q,k,p+q+k) = \Gamma^{(1)}_{\nu\tau}(2,p, p + k )-\Gamma^{(1)}_{\nu\tau}(2,p + q, k,  k + p + q )\\
{\rm or}&&Q^\mu \Big(M^{(1)}_{\mu\nu\tau}\Big)_{arrr} (p,p+k) = \Big(\Gamma^{(1)}_{\nu\tau}\Big)_{arr}(p, k+p)-\Big(\Gamma^{(1)}_{\nu\tau}\Big)_{arr}(p + q, k + p + q )\nonumber
\eea
The last set of numbers in the last line of (\ref{WIbare}) is (32,8,15). The corresponding ward identity is
\bea
&& Q^\mu C^{(1)}_{\mu\nu\tau} (32,p,q,k,s,p+q+k+s) = M^{(1)}_{\nu\tau}(8,p, k, s, k + p + s)-M^{(1)}_{\nu\tau}(15,p + q, k, s, k + p + q + s)\nonumber\\
{\rm or}&&Q^\mu \Big(C^{(1)}_{\mu\nu\tau}\Big)_{aaaaa} (p,q,k,s,p+q+k+s) = \Big(M^{(1)}_{\nu\tau}\Big)_{aaar}(p, k, s, k + p + s)-\Big(M^{(1)}_{\nu\tau}\Big)_{raaa}(p + q, k, s, k + p + q + s)\nonumber
\eea

\subsection{Full Vertices}

We verify that the ward identities derived above hold for the full 3-point and 4-point vertex functions. 

\subsubsection{2-point vertex function}

We give the results for the 2-point vertex function shown in Fig (\ref{Sigma}). These expressions will be needed to verify the ward identities for the 3-point vertex function.
\par\begin{figure}[H]
\begin{center}
\includegraphics[width=6cm]{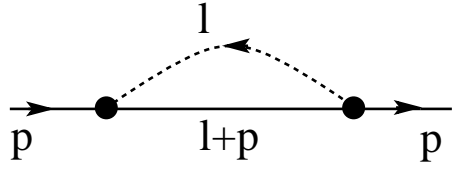}
\end{center}
\caption{The 2-point vertex function}
 \label{Sigma}
\end{figure}
 \bea
\label{PIres}
&& \Sigma(2,p) = -\frac{i }{2}\int \frac{d^4l}{(2\pi)^4}\left[\right.a(l) r(l+p) \Gamma (2,p,l+p) \left(\Gamma
(4,l+p,p)+\Gamma (3,l+p,p) N_F(l+p)-\Gamma (2,l+p,p) N_B(l)\right) \left.\right]\\
&& \Sigma(3,p) = -\frac{i }{2}\int \frac{d^4l}{(2\pi)^4}\left[\right.a(l+p) r(l) \Gamma (5,l+p,p) \left(\Gamma
(7,p,l+p)-\Gamma (3,p,l+p) N_F(l+p)+\Gamma (5,p,l+p) N_B(l)\right)  \left.\right]\nonumber
\\
&& \Sigma(4,p) =- \frac{i }{2}\int \frac{d^4l}{(2\pi)^4} \nonumber\\
&&~~\left[\right.a(l+p) r(l) \Gamma (5,l+p,p) \left(\Gamma (8,p,l+p)-\Gamma (4,p,l+p) N_F(l+p)+\left(\Gamma
(6,p,l+p)-\Gamma (2,p,l+p) N_F(l+p)\right)
   N_B(l)\right)\nonumber\\
&&~~~+a(l) r(l+p) \Gamma (2,p,l+p) \left(\Gamma (8,l+p,p)+\Gamma (7,l+p,p) N_F(l+p)-\left(\Gamma (6,l+p,p)+\Gamma (5,l+p,p)
   N_F(l+p)\right) N_B(l)\right)  \left.\right]\nonumber
\eea

\subsection{3-point vertex function}

First we calculate all seven components of the three graphs shown in Fig (\ref{Gamma}) that contribute to the 3-point vertex function. We give the result for one example: $\Gamma[2,p,p+q]$. 
\par\begin{figure}[H]
\begin{center}
\includegraphics[width=12cm]{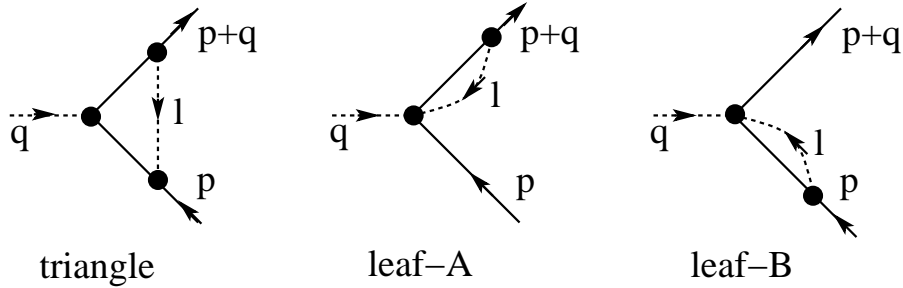}
\end{center}
\caption{The 3-point vertex function}
 \label{Gamma}
\end{figure}
\bea
\label{resG2}
&&\Gamma_{\rm triangle}(2,p,p+q) = \frac{i}{2}\int \frac{d^4l}{(2\pi)^4}\left[\right. 
a(l) r(l+p) \Gamma (2,p,l+p) \nonumber\\
&&~~(a(l+s) \Gamma (3,l+s,s) (\Gamma (6,l+p,l+s)+\Gamma (5,l+p,l+s) N_F(l+p)-\Gamma (2,l+p,l+s)
   N_F(l+s))\nonumber\\
&&~~+r(l+s) \Gamma (2,l+p,l+s) (\Gamma (4,l+s,s)+\Gamma (3,l+s,s) N_F(l+s)-\Gamma (2,l+s,s) N_B(l)))
\left.\right]\nonumber\\
&&\Gamma_{\rm leafA}(2,p,p+q) = -\frac{i}{2}\int \frac{d^4l}{(2\pi)^4}\nonumber\\
&&~~\left[\right. 
a(l) r(l+s) M(2,p,q,l,l+s) (\Gamma (4,l+s,s)+\Gamma (3,l+s,s) N_F(l+s)-\Gamma (2,l+s,s) N_B(l))
\left.\right]\nonumber\\
&&\Gamma_{\rm leafB}(2,p,p+q) = -\frac{i}{2}\int \frac{d^4l}{(2\pi)^4}\nonumber\\
&&~~\left[\right. 
a(l) r(l+p) \Gamma (2,p,l+p) (M(6,l+p,q,-l,s)+M(5,l+p,q,-l,s) N_F(l+p)-M(2,l+p,q,-l,s) N_B(l))
\left.\right]\nonumber
\eea

Next, we verify the ward identities for the seven vertex functions. We give detailed results for one example: $\Gamma[2,p,p+q]$. Starting from (\ref{resG2}) and contracting with $Q$ we obtain:
\bea
\label{GAMMAres}
Q\cdot \Gamma_{\rm triangle}(2,p,p+q) = && {\cal X}[2]+{\cal Y}[2] \\
Q\cdot \Gamma_{\rm leafA}(2,p,p+q) = && -{\cal X}[2]+{\bf x}[2] \nonumber\\
Q\cdot \Gamma_{\rm leafB}(2,p,p+q) = && -{\cal Y}[2]+{\bf y}[2] \nonumber
\eea
where
\bea
\label{GAMMAres2}
{\cal X}[2] &&= \frac{i}{2}\int \frac{d^4l}{(2\pi)^4}\left[\right. a(l) r(l+p+q)\Gamma (2,p,l+p)  \\
&&~~\left(\Gamma (4,l+p+q,p+q)+\Gamma (3,l+p+q,p+q) N_F(l+p+q)-\Gamma (2,l+p+q,p+q) N_B(l)\right)\left.\right]\nonumber\\
{\cal Y}[2] &&=-\frac{i}{2}\int \frac{d^4l}{(2\pi)^4}\left[\right.a(l) r(l+p)\Gamma (2,p,l+p)\nonumber\\
&&~~  \left(\Gamma (4,l+p+q,p+q)+\Gamma (3,l+p+q,p+q) N_F(l+p)-\Gamma (2,l+p+q,p+q) N_B(l)\right)\left.\right]\nonumber\\
{\bf x}[2] &&=\frac{i}{2}\int \frac{d^4l}{(2\pi)^4}\left[\right. a(l) r(l+p+q) \Gamma (2,p+q,l+p+q)\nonumber\\
&&~~ \left(\Gamma (4,l+p+q,p+q)+\Gamma (3,l+p+q,p+q) N_F(l+p+q)-\Gamma (2,l+p+q,p+q) N_B(l)\right)\left.\right]\nonumber\\
{\bf y}[2] &&=-\frac{i}{2}\int \frac{d^4l}{(2\pi)^4}\left[\right.
a(l) r(l+p) \Gamma (2,p,l+p)\nonumber\\
&&~~ \left(\Gamma (4,l+p,p)+\Gamma (3,l+p,p) N_F(l+p)-\Gamma (2,l+p,p) N_B(l)\right)\left.\right]\nonumber
\eea
Comparing with (\ref{PIres}) we obtain
\bea
Q\cdot \Gamma(2,p,p+q) = Q\cdot \Big(\Gamma_{\rm triangle}(2,p,p+q)+\Gamma_{\rm leafA}(2,p,p+q)+\Gamma_{\rm leafB}(2,p,p+q)\Big) = \Sigma(2,p)-\Sigma(2,p+q)\nonumber
\eea
which agrees with (\ref{example1}). The results for all components agree with the results listed in (\ref{WIbare}).

\subsubsection{4-point vertex function}

We verify the ward identity for the 15 4-point vertex functions. We give detailed results for one example: $M(2,p,q,k,p+q+k)$. There are five types of diagrams to consider. They are shown in Figs. (\ref{Box}), (\ref{LeafTail}), (\ref{JellyFish}), (\ref{SeaGull}), (\ref{PolyWog}). 
\par\begin{figure}[H]
\begin{center}
\includegraphics[width=8cm]{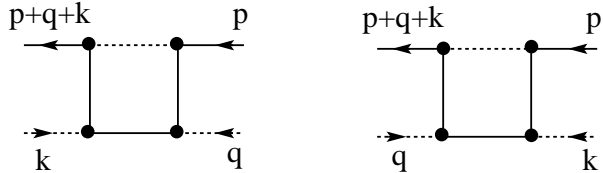}
\end{center}
\caption{The box and crossed-box diagrams}
 \label{Box}
\end{figure}
\par\begin{figure}[H]
\begin{center}
\includegraphics[width=12cm]{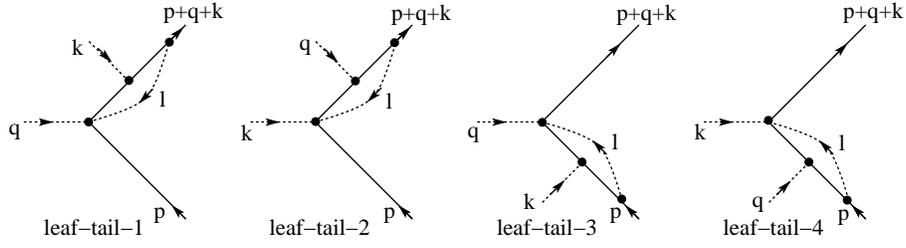}
\end{center}
\caption{The leaftail diagrams}
 \label{LeafTail}
\end{figure}
\par\begin{figure}[H]

\begin{center}
\includegraphics[width=8cm]{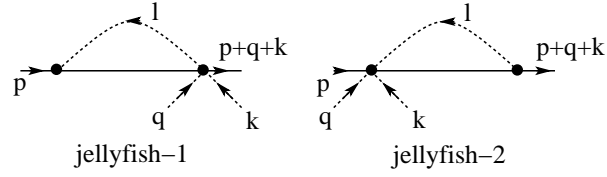}
\end{center}
\caption{The jellyfish diagrams}
 \label{JellyFish}
\end{figure}
\par\begin{figure}[H]
\begin{center}
\includegraphics[width=10cm]{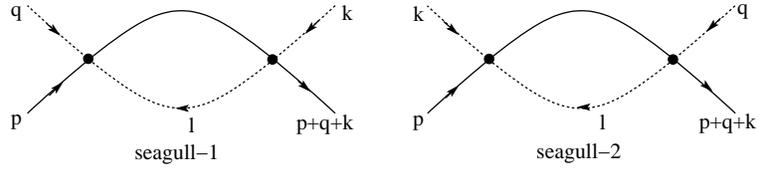}
\end{center}
\caption{The seagull diagrams}
 \label{SeaGull}
\end{figure}
\par\begin{figure}[H]
\begin{center}
\includegraphics[width=4cm]{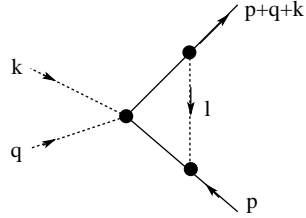}
\end{center}
\caption{The polywog diagram}
 \label{PolyWog}
\end{figure}

Contracting with $Q$ and using (\ref{GAMMAres}) and (\ref{GAMMAres2}) we obtain:
\bea
&&\Gamma_{\rm seagull-1}(2,p,q,k,u) =  \alpha[2]+A[2]\,;~~\Gamma_{\rm seagull-2}(2,p,q,k,u) = \beta[2] + B[2]\\
&& \Gamma_{\rm leaftail-1}(2,p,q,k,u) = \gamma[2] + C[2]\,;~~\Gamma_{\rm leaftail-3}(2,p,q,k,u) = \delta[2] + D[2]\nonumber\\
&& \Gamma_{\rm jellyfish-1}(2,p,q,k,u) = \epsilon[2] + E[2]\,;~~\Gamma_{\rm jellyfish-2}(2,p,q,k,u) = \phi[2] + F[2]\nonumber\\[2mm]
&& \Gamma_{\rm polywog}(2,p,q,k,u) = G[2]+H[2] \nonumber\\[2mm]
&& \Gamma_{\rm leaftail-2}(2,p,q,k,u) = -B[2]-F[2]\,;~~\Gamma_{\rm leaftail-4}(2,p,q,k,u) = -A[2]-E[2]\nonumber\\
&& \Gamma_{\rm box}(2,p,q,k,u) = -G[2]-C[2] \,;~~\Gamma_{\rm crossed-box}(2,p,q,k,u) = -H[2]-D[2] \nonumber
\eea
where 
\bea
&&\alpha[2] = -\Gamma_{\rm leafB}(2,p+q,u) \\
&&\beta[2] = \Gamma_{\rm leafA}(2,p,t)\nonumber \\
&&\gamma[2] = -\Gamma_{\rm triangle}(2,p+q,u)\nonumber \\
&&\delta[2] = \Gamma_{\rm triangle}(2,p,t)\nonumber \\
&&\epsilon[2] = \Gamma_{\rm leafB}(2,p,t)\nonumber \\
&&\phi[2] = -\Gamma_{\rm leafA}(2,p+q,u)\nonumber \\
&&A[2] = -\frac{i}{2}\int \frac{d^4l}{(2\pi)^4}\nonumber\\
&&~~\left[\right. 
a(l) r(l+s) \Gamma (2,p,l+p) (M(6,l+s,k,-l,u)+M(3,l+s,k,-l,u) N_F(l+s)-M(1,l+s,k,-l,u) N_B(l))
\left.\right]\nonumber\\
&&B[2] = \frac{i}{2}\int \frac{d^4l}{(2\pi)^4}\left[\right. 
a(l) r(l+t) M(2,p,k,l,l+t) (\Gamma (4,l+u,u)+\Gamma (3,l+u,u) N_F(l+t)-\Gamma (2,l+u,u) N_B(l))
\left.\right]\nonumber\\
&&C[2] = -\frac{i}{2}\int \frac{d^4l}{(2\pi)^4}\left[\right. 
a(l) r(l+s) \Gamma (2,p,l+p)\nonumber\\
&&~~ (a(l+u) \Gamma (3,l+u,u) (\Gamma (6,l+s,l+u)+\Gamma (5,l+s,l+u) N_F(l+s)-\Gamma (2,l+s,l+u)
   N_F(l+u))\nonumber\\
&&~~+r(l+u) \Gamma (2,l+s,l+u) (\Gamma (4,l+u,u)+\Gamma (3,l+u,u) N_F(l+u)-\Gamma (2,l+u,u) N_B(l)))
\left.\right]\nonumber\\
&&D[2] = \frac{i}{2}\int \frac{d^4l}{(2\pi)^4}\left[\right. 
a(l) r(l+p) \Gamma (2,p,l+p)\nonumber\\
&&~~ (a(l+t) \Gamma (3,l+u,u) (\Gamma (6,l+p,l+t)+\Gamma (5,l+p,l+t) N_F(l+p)-\Gamma (2,l+p,l+t)
   N_F(l+t))\nonumber\\
&&~~r(l+t) \Gamma (2,l+p,l+t) (\Gamma (4,l+u,u)+\Gamma (3,l+u,u) N_F(l+t)-\Gamma (2,l+u,u) N_B(l)))
\left.\right]\nonumber\\
&&E[2] = \frac{i}{2}\int \frac{d^4l}{(2\pi)^4}\nonumber\\
&&~~\left[\right. 
a(l) r(l+p) \Gamma (2,p,l+p) (M(6,l+s,k,-l,u)+M(5,l+s,k,-l,u) N_F(l+p)-M(2,l+s,k,-l,u) N_B(l))
\left.\right]\nonumber\\
&&F[2] = -\frac{i}{2}\int \frac{d^4l}{(2\pi)^4}\left[\right. 
a(l) r(l+u) M(2)(p,k,l,l+t) (\Gamma (4,l+u,u)+\Gamma (3,l+u,u) N_F(l+u)-\Gamma (2,l+u,u) N_B(l))
\left.\right]\nonumber\\
&&G[2] = \frac{i}{2}\int \frac{d^4l}{(2\pi)^4}\left[\right. 
a(l) r(l+p) \Gamma (2,p,l+p) \nonumber\\
&&~~(a(l+u) \Gamma (3,l+u,u) (\Gamma (6,l+s,l+u)+\Gamma (5,l+s,l+u) N_F(l+p)-\Gamma (2,l+s,l+u)
   N_F(l+u))\nonumber\\
&&~~+r(l+u) \Gamma (2,l+s,l+u) (\Gamma (4,l+u,u)+\Gamma (3,l+u,u) N_F(l+u)-\Gamma (2,l+u,u) N_B(l)))
\left.\right]\nonumber\\
&&H[2] = -\frac{i}{2}\int \frac{d^4l}{(2\pi)^4}\left[\right. 
a(l) r(l+p) \Gamma (2,p,l+p) \nonumber\\
&&~~(a(l+u) \Gamma (3,l+u,u) (\Gamma (6,l+p,l+t)+\Gamma (5,l+p,l+t) N_F(l+p)-\Gamma (2,l+p,l+t)
   N_F(l+u))\nonumber\\
&&~~+r(l+u) \Gamma (2,l+p,l+t) (\Gamma (4,l+u,u)+\Gamma (3,l+u,u) N_F(l+u)-\Gamma (2,l+u,u) N_B(l)))
\left.\right]\,.\nonumber
\eea
Using $\Gamma = \Gamma_{\rm triangle}+\Gamma_{\rm leafA}+\Gamma_{\rm leafB}$ and combining we have
\bea
Q^\mu\cdot M_{\mu\nu}(2,p,q,k,u) =  \Gamma_\nu(2,p,t)-\Gamma_\nu(2,p+q,u)\nonumber
\eea
which agrees with (\ref{example2}). The results for all components agree with the results listed in (\ref{WIbare}).

\section{Conclusions}
\label{conc}
Calculations in real time statistical field theory are complicated by the extra indices that result from the doubling of degrees of 
freedom. Because of this technical problem, many people avoid the real time formalism of finite temperature field theory in spite 
of its significant advantages, as compared with the imaginary time formalism. Two of the major advantages of working in real time 
are the fact that  analytic continuations are not necessary, and that it is easy to generalize to non-equilibrium situations. 
In this paper we have made a contribution towards reducing the technical difficulties associated with the real time formulation 
of statistical field theory. 
We have written a {\it Mathematica} program that performs the contractions over the tensor indices that appear in real
time statistical field theory and determines the
integrand corresponding to any amplitude. 
The program is designed so that it can be used by someone with no 
previous experience with {\it Mathematica}. It is available on the 
internet at
www.brandonu.ca/physics/fugleberg/Research/Dick.html.
It can be used in the 1-2,
Keldysh or RA basis, and it can do calculations in or out of equilibrium.

We have used the program to calculate the QED ward
identity for the 3-point function ($2^n-1 = 7$ components), the 4-point function for two fermions and two photons (($2^n-1 = 15$ components),
and the 5-point function for two fermions and three photons ($2^n-1 = 31$ components).  Some of these identities have appeared previously in
the literature, but the complete set of identities has not previously been published. The calculation therefore serves two purposes: 
it provides a check of the program, and it produces useful new information. 
We give a table that lists the
results for the ward identities (Eqn. (\ref{WIbare})).  In addition, we give a simple
general expression for the KMS conditions between $n$-point 
functions and vertex functions, in both the Keldysh and RA bases (Eqns. (\ref{general_KMS_physical}) and (\ref{general_KMS_RA})).

\end{document}